\documentclass[english]{article}
\usepackage[T1]{fontenc}
\usepackage[latin9]{inputenc}
\usepackage{geometry}
\usepackage[T1]{fontenc}
\usepackage[latin9]{inputenc}
\usepackage{color}
\usepackage{array}
\usepackage{amsmath}
\usepackage{graphicx}
\usepackage{amssymb}
\usepackage{epstopdf}
\usepackage{caption}
\usepackage{epsfig}
\usepackage{bm}
\usepackage{epsfig}
\usepackage{color}
\usepackage{subfig}
\usepackage{babel}

\makeatletter

\DeclareRobustCommand{\greektext}{%
  \fontencoding{LGR}\selectfont\def\encodingdefault{LGR}}
\DeclareRobustCommand{\textgreek}[1]{\leavevmode{\greektext #1}}
\DeclareFontEncoding{LGR}{}{}
\DeclareTextSymbol{\~}{LGR}{126}
\newcommand{\lyxmathsym}[1]{\ifmmode\begingroup\def\b@ld{bold}
  \text{\ifx\math@version\b@ld\bfseries\fi#1}\endgroup\else#1\fi}

\providecommand{\tabularnewline}{\\}

\@ifundefined{showcaptionsetup}{}{%
 \PassOptionsToPackage{caption=false}{subfig}}
\usepackage{subfig}
\makeatother

\usepackage{babel}

\begin{document}

\title{Nonclassical properties of a contradirectional nonlinear optical
coupler}

\maketitle
\begin{center}
Kishore Thapliyal$^{a}$, Anirban Pathak$^{a,b,}$%
\footnote{Email: anirban.pathak@gmail.com, 

Phone: +91 9717066494%
}, Biswajit Sen$^{c}$, and Jan ${\rm \check{Perina}}$$^{b,d}$ 
\par\end{center}

\begin{center}
$^{a}$Jaypee Institute of Information Technology, A-10, Sector-62,
Noida, UP-201307, India 
\par\end{center}

\begin{center}
$^{b}$RCPTM, Joint Laboratory of Optics of Palacky University and
Institute of Physics of Academy of Science of the Czech Republic,
Faculty of Science, Palacky University, 17. listopadu 12, 771 46 Olomouc,
Czech Republic
\par\end{center}

\begin{center}
$^{c}$Department of Physics, Vidyasagar Teachers' Training College,
Midnapore-721101, India
\par\end{center}

\begin{center}
$^{d}$Department of Optics, Palacky University, 17. listopadu 12,
771 46 Olomouc, Czech Republic
\par\end{center}
\begin{abstract}
We investigate the nonclassical properties of output fields propagated
through a contradirectional asymmetric nonlinear optical coupler consisting
of a linear waveguide and a nonlinear (quadratic) waveguide operated
by second harmonic generation. In contrast to the earlier results,
all the initial fields are considered weak and a completely quantum
mechanical model is used here to describe the system. Perturbative
solutions of Heisenberg's equations of motion for various field modes
are obtained using Sen-Mandal technique. Obtained solutions are subsequently
used to show the existence of\textcolor{red}{{} }single-mode and intermodal
squeezing, single-mode and intermodal antibunching, two-mode and multi-mode
entanglement in the output of contradirectional asymmetric nonlinear
optical coupler. Further, existence of higher order nonclassicality
is also established by showing the existence of higher order antibunching,
higher order squeezing and higher order entanglement. Variation of
observed nonclassical characters with different coupling constants
and phase mismatch is discussed.
\end{abstract}

\section{Introduction}

Different aspects of nonclassical properties of electromagnetic field
have been studied since the advent of quantum optics. However, the
interest on nonclassical states has been considerably escalated with
the progress of interdisciplinary field of quantum computation and
quantum communication in recent past as a large number of applications
of nonclassical states have been reported in context of quantum computation
and quantum communication \cite{CV-qkd-hillery,teleportation of coherent state,antibunching-sps,Ekert protocol,Bennet1993,densecoding}.
Specifically, it is shown that squeezed states can be used for the
implementation of continuous variable quantum cryptography \cite{CV-qkd-hillery}
and teleportation of coherent states \cite{teleportation of coherent state},
antibunched states can be used to build single photon sources \cite{antibunching-sps},
entangled states are essential for the implementation of a set of
protocols of discrete \cite{Ekert protocol} and continuous variable
quantum cryptography \cite{CV-qkd-hillery}, quantum teleportation
\cite{Bennet1993}, dense-coding \cite{densecoding}, etc., states
violating Bell's inequality are useful for the implementation of protocols
of device independent quantum key distribution \cite{device ind}.
Study of the possibility of generation of nonclassical states (specially,
entanglement and nonlocal characters of quantum states) in different
quantum systems have recently become extremely relevant and important
for the researchers working in different aspects of quantum information
theory and quantum optics. Existence of entanglement and other nonclassical
states in a large number of bosonic systems have already been reported
(See \cite{pathak-perina,pathak-PRA,pathak-pra2} and references therein).
However, it is still interesting to study the possibility of generation
of nonclassical states in experimentally realizable simple systems.
A specific system of this kind is a nonlinear optical coupler which
can be easily realized using optical fibers or photonic crystals.
Optical couplers are interesting for several reasons. Firstly, in
a coupler the amount of nonclassicality can be controlled by controlling
the interaction length and the coupling constants. Further, optical
couplers are of specific interest as recently Matthews et al. have
experimentally demonstrated manipulation of multiphoton entanglement
in quantum circuits constructed using optical couplers \cite{Nature09},
Mandal and Midda have shown that NAND gate (thus, in principle a classical
computer) can be build using nonlinear optical couplers \cite{Mandal-Midda}.
Motivated by these facts we aim to systematically investigate the
possibility of observing nonclassicality in nonlinear optical couplers.
As a first effort we have reported lower order and higher order entanglement
and other higher order nonclassical effects in an asymmetric codirectional
nonlinear optical coupler which is prepared by combining a linear
waveguide and a nonlinear (quadratic) waveguide operated by second
harmonic generation \cite{kishore-1}. Waveguides interact with each
other through evanescent wave. Extending the earlier investigation
\cite{kishore-1} here we study a similar asymmetric nonlinear optical
coupler for contradirectional propagation of fields. This type of
contradirectional asymmetric optical coupler was studied earlier by
some of us \cite{perina-review,contra-1}. However, in the earlier
studies \cite{perina-review,contra-1} intermodal entanglement and
some of the higher order nonclassical properties studied here were
not studied. Further, in those early studies second-harmonic mode
($b_{2})$ was assumed to be pumped with a strong coherent beam. In
other words, $b_{2}$ mode was assumed to be classical and thus it
was beyond the scope of the previous studies to investigate single-mode
and intermodal nonclassicalities involving this mode. Interestingly,
completely quantum mechanical treatment adopted in the present work
is found to show intermodal squeezing in a compound mode involving
$b_{2}$ mode. In addition, conventional short-length solutions of
Heisenberg's equations of motion were used in earlier studies \cite{perina-review,contra-1},
but recently it is established by some of us that using improved perturbative
solutions obtained by Sen-Mandal approach \cite{bsen1,Bsen-2,bsen3}
we can observe several nonclassical characters that are not observed
using short-length (or short-time approach) \cite{pathak-PRA,pathak-pra2,kishore-1,Mandal-Perina}.
Keeping these facts in mind present paper aims to study nonclassical
properties of this contradirectional asymmetric nonlinear optical
coupler with specific attention to entanglement using perturbative
solution obtained by Sen-Mandal technique.

Nonclassical properties of different optical couplers are extensively
studied in past (see \cite{perina-review} for a review). Specially,
signatures of nonclassicality in terms of photon statistics, phase
properties and squeezing were investigated in codirectional and contradirectional
Kerr nonlinear couplers having fixed and varying linear coupling constant
\cite{Kerr-1,kerr-2,Kerr-3,kerr-4,kerr-5}, Raman and Brillouin coupler
\cite{Raman and Brilloin} and parametric coupler \cite{parametric,parametric2},
asymmetric \cite{contra-1,Mandal-Perina,assymetric-strong-pump,assymetric-2,nonclassical-input1}
and symmetric \cite{nonclassical-input1,co-and-contra,nonclasssical input2}
directional nonlinear coupler etc. Here it would be apt to note that
the specific coupler system that we wish to study in the present paper
has already been investigated \cite{contra-1,co and contra with phase mismatch}
for contradirectional propagation of classical (coherent) input modes
and it is shown that depth of nonclassicality may be controlled by
varying the phase mismatching $\Delta k$ \cite{co and contra with phase mismatch}.

Existing studies are restricted to the investigation of lower order
nonclassical effects (e.g., squeezing and antibunching) under the
conventional short-length approximation. Only a few discrete efforts
have recently been made to study higher order nonclassical effects
and entanglement in optical couplers \cite{kishore-1,leonoski1,thermal ent-kerr,kerr-lionoski-miranowicz,GBS,higher-order,adam},
but except a recent study on codirectional nonlinear optical coupler
reported by us \cite{kishore-1}, all the other efforts were limited
to Kerr nonlinear optical coupler. For example, in 2004, Leonski and
Miranowicz reported entanglement in Kerr nonlinear optical coupler
\cite{kerr-lionoski-miranowicz} and pumped Kerr nonlinear optical
coupler \cite{adam}, subsequently entanglement sudden death \cite{leonoski1}
and thermally induced entanglement \cite{thermal ent-kerr} were reported
in the same system. Amplitude squared (higher order) squeezing was
also reported in Kerr nonlinear optical coupler \cite{higher-order}.
However, no effort has yet been made to rigorously study the higher
order nonclassical effects and entanglement in contradirectional nonlinear
optical coupler. Keeping these facts in mind in the present letter
we aim to study nonclassical effects (including higher order nonclassicality
and entanglement) in contradirectional nonlinear optical coupler.

Remaining part of the paper is organized as follows. In Section \ref{sec:The-model-and},
the model momentum operator that  represents the asymmetric nonlinear
optical coupler is described and perturbative solutions of equations
of motion corresponding to different field modes present in the momentum
operator are reported. In Section \ref{sec:Criteria-of-nonclassicalities},
we briefly describe a set of criteria of nonclassicality. In Section
\ref{sec:Nonclassicality-in-contradirectional} the criteria described
in the previous section are used to investigate the existence of different
nonclassical characters (e.g., lower order and higher order squeezing,
antibunching, and entanglement) in various field modes present in
the contradirectional asymmetric nonlinear optical coupler. Finally,
Section \ref{sec:Conclusions} is dedicated for conclusions.

\section{{\normalsize The model and the solution\label{sec:The-model-and}}}

\begin{figure}
\centering{}\includegraphics[scale=0.8]{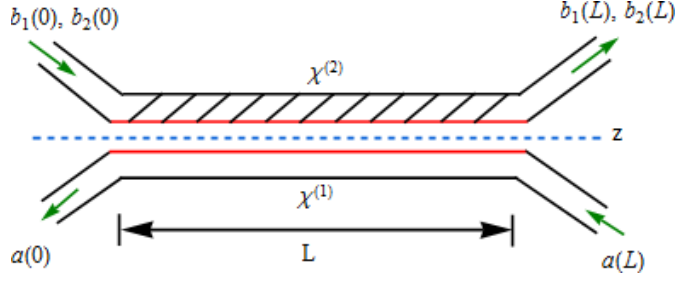}\caption{\textcolor{blue}{\label{fig:schematic diagram}}(Color online) Schematic
diagram of a contradirectional asymmetric nonlinear optical coupler
prepared by combining a linear wave guide ($\chi^{(1)}$) with a nonlinear
($\chi^{(2)}$) waveguide operated by second harmonic generation.
The fields involved are described by the corresponding annihilation
operators, as shown; $L$ is the interaction length. }
\end{figure}

A schematic diagram of a contradirectional asymmetric nonlinear optical
coupler is shown in Fig. \ref{fig:schematic diagram}. From Fig. \ref{fig:schematic diagram}
one can easily observe that a linear waveguide is combined with a
nonlinear $\left(\chi^{(2)}\right)$ waveguide to constitute the asymmetric
coupler of our interest. Further, from Fig. \ref{fig:schematic diagram}
we can observe that in the linear waveguide field propagates in a
direction opposite to the propagation direction of the field in the
nonlinear waveguide. Electromagnetic field characterized by the bosonic
field annihilation (creation) operator $a\,(a^{\dagger})$ propagates
through the linear waveguide. Similarly, the field operators $b_{i}\,(b_{i}^{\dagger})$
corresponds to the nonlinear medium. Specifically, $b_{1}(k_{1})$
and $b_{2}(k_{2})$ denote annihilation operators (wave vectors) for
fundamental and second harmonic modes, respectively. Now the momentum
operator for contradirectional optical coupler is \cite{perina-review}\begin{equation}
G=-\hbar kab_{1}^{\dagger}-\hbar\Gamma b_{1}^{2}b_{2}^{\dagger}\exp(i\Delta kz)\,+{\rm h.c}.\label{eq:1}\end{equation}
where ${\rm h.c.}$ stands for the Hermitian conjugate and $\Delta k=|2k_{1}-k_{2}|$
represents the phase mismatch between the fundamental and second harmonic
beams. The linear (nonlinear) coupling constant, proportional to susceptibility
$\chi^{(1)}$ $(\chi^{(2)})$, is denoted by the parameter $k$ $\left(\Gamma\right)$.
It is reasonable to assume $\chi^{(2)}\ll\chi^{(1)}$ as in a real
physical system we usually obtain $\chi^{(2)}/\chi^{(1)}\,\simeq10^{-6}$.
As a consequence, in absence of a highly strong pump $\Gamma\ll k$.
Earlier this model of contradirectional optical coupler was investigated
by some of the present authors (\cite{perina-review} and references
therein). Using (\ref{eq:1}) and the procedure described in \cite{perina-review}
we can obtain the coupled differential equations for three different
modes as follows \begin{equation}
\frac{da}{dz}=ik^{*}b_{1},\,\frac{db_{1}}{dz}=-ika-2i\Gamma^{*}b_{1}^{\dagger}b_{2}\exp\left(-i\Delta kz\right),\,\frac{db_{2}}{dz}=-i\Gamma b_{1}^{2}\exp\left(i\Delta kz\right).\label{eq:aaa}\end{equation}
Here it would be apt to mention that momentum operator for contradirectional
asymmetric nonlinear coupler (\ref{eq:1}) is same as that of codirectional
asymmetric nonlinear optical coupler \cite{perina-review}. However,
for the contradirectional couplers the sign of derivative in the Heisenberg's
equation of motion of the contra-propagating mode is changed (i.e.,
in the present case $\frac{da}{dz}$ is replaced by $-\frac{da}{dz}$
as mode $a$ is considered here as the contra-propagating mode). The
method used here to obtain (\ref{eq:aaa}) is described in Refs. \cite{perina-review,contra-1}.
Further, this particular description of contradirectional coupler
is valid only for the situation when the forward propagating waves
reach $z=L$ and the counter (backward) propagating wave reach $z=0$.
Thus the coupled equations described by (\ref{eq:aaa}) and their
solution obtained below are not valid for $0<z<L$ \cite{contra-1}.
Earlier these coupled equations (\ref{eq:aaa}) were solved under
short-length approximation. Here we aim to obtain perturbative solution
for these equations using Sen-Mandal method which is already shown
to be useful in detecting nonclassical characters not identified by
short-length solution \cite{kishore-1,Mandal-Perina}. Keeping this
in mind, we plan to solve (\ref{eq:aaa}) using Sen-Mandal approach.

Using Sen-Mandal approach we have obtained closed form perturbative
analytic solutions of (\ref{eq:aaa}) as\begin{equation}
\begin{array}{lcl}
a(0) & = & f_{1}a(L)+f_{2}b_{1}(0)+f_{3}b_{1}^{\dagger}(0)b_{2}(0)+f_{4}a^{\dagger}(L)b_{2}(0),\\
b_{1}(L) & = & g_{1}a(L)+g_{2}b_{1}(0)+g_{3}b_{1}^{\dagger}(0)b_{2}(0)+g_{4}a^{\dagger}(L)b_{2}(0),\\
b_{2}(L) & = & h_{1}b_{2}(0)+h_{2}b_{1}^{2}(0)+h_{3}b_{1}(0)a(L)+h_{4}a^{2}(L),\end{array}\label{2}\end{equation}
with\begin{equation}
\begin{array}{lcl}
f_{1} & = & g_{2}={\rm sech}|k|L,\\
f_{2} & = & -g_{1}^{*}=-\frac{ik^{*}\tanh|k|L}{|k|},\\
f_{3} & = & Ck^{*}\Delta kf_{1}^{2}\left\{ i\Delta k\sinh2|k|L+2|k|\left(G_{+}-1-\cosh2|k|L\right)\right\} ,\\
f_{4} & = & 2Ck^{*2}f_{1}^{2}\left\{ i\Delta k\sinh|k|LG_{+}-2|k|\cosh|k|LG_{-}\right\} ,\\
g_{3} & = & -2C|k|f_{1}^{2}\left\{ \left(\Delta k^{2}+2|k|^{2}\right)\cosh|k|LG_{-}+i\Delta k|k|\sinh|k|LG_{+}\right\} ,\\
g_{4} & = & Ck^{*}\Delta kf_{1}^{2}\left\{ i\Delta k\sinh2|k|L\left(G_{+}-1\right)-2|k|\left(1-\cosh2|k|L\left(G_{+}-1\right)\right)\right\} ,\\
h_{1} & = & 1,\\
h_{2} & = & \frac{C^{*}|k|}{2}f_{1}^{2}\left\{ 4|k|^{2}G_{-}^{*}+\Delta k^{2}\left(1-2\left(G_{+}^{*}-1\right)+\cosh2|k|L\right)-2i\Delta k|k|\sinh2|k|L\right\} ,\\
h_{3} & = & 2C^{*}kf_{1}^{2}\left\{ \Delta k|k|G_{-}^{*}\cosh|k|L+\left[i\Delta k^{2}-2i|k|^{2}G_{-}^{*}\right]\sinh|k|L\right\} ,\\
h_{4} & = & \frac{C^{*}|k|k}{k^{*}}f_{1}\left\{ 2|k|^{2}f_{1}G_{-}^{*}+\Delta k\sinh|k|L\left(G_{+}^{*}-1\right)\left[2i|k|+\Delta k\tanh|k|L\right]\right\} ,\end{array}\label{eq:soln}\end{equation}
where $C=\frac{\Gamma^{*}}{|k|\Delta k\left(\Delta k^{2}+4|k|^{2}\right)}$
and $G_{\pm}=\left(1\pm\exp(-i\Delta kL)\right)$. The solution obtained
above is verified by ESCR (Equal Space Commutation Relation) which
implies $\left[a\left(0\right),a^{\dagger}\left(0\right)\right]=\left[b_{1}\left(L\right),b_{1}^{\dagger}\left(L\right)\right]=\left[b_{2}\left(L\right),b_{2}^{\dagger}\left(L\right)\right]=1$
while all other equal space commutations are zero. Further, we have
verified that the solutions reported here satisfies constant of motion.
To be precise, in Ref. \cite{contra-1}, it was shown that the constant
of motion for the present system leads to \begin{equation}
a^{\dagger}\left(0\right)a\left(0\right)+b_{1}^{\dagger}\left(L\right)b_{1}\left(L\right)+2b_{2}^{\dagger}\left(L\right)b_{2}\left(L\right)=a^{\dagger}\left(L\right)a\left(L\right)+b_{1}^{\dagger}\left(0\right)b_{1}\left(0\right)+2b_{2}^{\dagger}\left(0\right)b_{2}\left(0\right).\label{eq:const-of-motion}\end{equation}
Here we have verified that the solution proposed here satisfies (\ref{eq:const-of-motion}).
We have also verified that the solution of contradirectional coupler
using short-length solution method reported in \cite{perina-review}
can be obtained as a special case of the present solution. Specifically,
to obtain the short-length solution we need to expand the trigonometric
functions present in the above solution and neglect all the terms
beyond quadratic powers of $L$ and consider phase mismatch $\Delta k=0$.
After doing so we obtain\begin{equation}
\begin{array}{c}
f_{1}=g_{2}=(1-\frac{1}{2}\left|k\right|^{2}L^{2}),\, f_{2}=-g_{1}^{*}=-ik^{*}L,\, f_{3}=-g_{4}=-\Gamma^{*}k^{*}L^{2},\\
g_{3}=-2i\Gamma^{*}L,\, h_{2}=-i\Gamma L,\, h_{3}=-\Gamma kL^{2}{\rm \, and\,}\, h_{1}=1,\, f_{4}=h_{4}=0,\end{array}\label{eq:shortlength}\end{equation}
which coincides with the short-length solution reported earlier by
some of the present authors \cite{perina-review,contra-1}. Clearly
the solution obtained here is valid and more general than the conventional
short-length solution as the solution reported here is fully quantum
solution and is valid for any length, restricting the coupling constant
only. Further, in case of codirectional optical coupler we have already
seen that several nonclassical phenomena not identified by short-length
solution are identified by the perturbative solution obtained by Sen-Mandal
method \cite{kishore-1,Mandal-Perina}. Keeping this in mind, in what
follows we investigate nonclassical characters of the fields that
have propagated through a contradirectional asymmetric nonlinear optical
coupler.

\section{{\normalsize Criteria of nonclassicality\label{sec:Criteria-of-nonclassicalities}}}

As the criteria for obtaining signatures of different nonclassical
phenomena are expressed in terms of expectation values of functions
of annihilation and creation operators of various modes, we can safely
state that Eqs. (\ref{2}) and (\ref{eq:soln}) provide us sufficient
resource for the investigation of the nonclassical phenomena. To illustrate
this point we may note that the criteria for quadrature squeezing
in single-mode ($j)$ and compound mode ($j,l)$ are \cite{Loudon}
\begin{equation}
\left(\Delta X_{j}\right)^{2}<\frac{1}{4}\,\mathrm{or}\,\left(\Delta Y_{j}\right)^{2}<\frac{1}{4}\label{eq:condition for squeezing}\end{equation}
and \begin{equation}
\left(\Delta X_{jl:j\neq l}\right)^{2}<\frac{1}{4}\,\mathrm{or}\,\left(\Delta Y_{jl:j\neq l}\right)^{2}<\frac{1}{4},\label{eq:condition for squeezing-1}\end{equation}
where $j,l\in\{a,\, b_{1},\, b_{2}\}$ and the quadrature operators
are defined as \begin{equation}
\begin{array}{lcl}
X_{a} & = & \frac{1}{2}\left(a+a^{\dagger}\right),\\
Y_{a} & = & -\frac{i}{2}\left(a-a^{\dagger}\right),\end{array}\label{eq:quadrature}\end{equation}
 and \begin{equation}
\begin{array}{lcl}
X_{ab} & = & \frac{1}{2\sqrt{2}}\left(a+a^{\dagger}+b+b^{\dagger}\right),\\
Y_{ab} & = & -\frac{i}{2\sqrt{2}}\left(a-a^{\dagger}+b-b^{\dagger}\right).\end{array}\label{eq:two-mode quadrature}\end{equation}
Similarly, the existence of single- and multi-mode nonclassical (sub-Poissonian)
photon statistics can be obtained through the following inequalities\begin{equation}
D_{a}=\left(\Delta N_{a}\right)^{2}-\left\langle N_{a}\right\rangle <0,\label{eq:antib-1}\end{equation}
 and \begin{equation}
D_{ab}=(\Delta N_{ab})^{2}=\left\langle a^{\dagger}b^{\dagger}ba\right\rangle -\left\langle a^{\dagger}a\right\rangle \left\langle b^{\dagger}b\right\rangle <0,\label{eq:antib2}\end{equation}
where (\ref{eq:antib-1}) provides us the condition for single-mode
antibunching%
\footnote{To be precise, this zero-shift correlation is more connected to sub-Poissonian
behavior. However, it is often referred to as antibunching \cite{Adam-criterion}.%
} and (\ref{eq:antib2}) provides us the condition for intermodal antibunching.
Many of the early investigations on nonclassicality were limited to
the study of squeezing and antibunching only, but with the recent
development of quantum computing and quantum information it has become
very relevant to study entanglement. Interestingly, there exist a
large number of inseparability criteria (\cite{NJP-GS-Ashoka,Adam-criterion}
and references therein) that can be expressed in terms of expectation
values of moments of field operators. For example, Hillery-Zubairy
criterion I and II (HZ-I and HZ-II) \cite{HZ-PRL,HZ2007,HZ2010} are
described as \begin{equation}
\begin{array}{lcl}
\left\langle N_{a}N_{b}\right\rangle  & -\left|\left\langle ab^{\dagger}\right\rangle \right|^{2}< & 0,\end{array}\label{hz1}\end{equation}
and \begin{equation}
\begin{array}{lcl}
\left\langle N_{a}\right\rangle \left\langle N_{b}\right\rangle  & -\left|\left\langle ab\right\rangle \right|^{2}< & 0,\end{array}\label{hz2}\end{equation}
respectively. Another, interesting criterion of inseparability that
can be expressed in terms of moments of creation and annihilation
operators is Duan et al.'s criterion which is described as follows
\cite{duan}: \begin{equation}
\begin{array}{lcl}
d_{ab}=\left(\Delta u_{ab}\right)^{2}+\left(\Delta v_{ab}\right)^{2}-2 & < & 0,\end{array}\label{duan-1}\end{equation}
 where \begin{equation}
\begin{array}{lcl}
u_{ab} & = & \frac{1}{\sqrt{2}}\left\{ \left(a+a^{\dagger}\right)+\left(b+b^{\dagger}\right)\right\} ,\\
v_{ab} & = & -\frac{i}{\sqrt{2}}\left\{ \left(a-a^{\dagger}\right)+\left(b-b^{\dagger}\right)\right\} .\end{array}\label{eq:duan-2-1}\end{equation}
Clearly our analytic solution (\ref{2})-(\ref{eq:soln}) enables
us to investigate intermodal entanglement using these criteria and
a bunch of other criteria of nonclassicality that are described in
Ref. \cite{Adam-criterion}. Interestingly, all the inseparability
criteria described above and in the remaining part of the present
work can be viewed as special cases of Shchukin-Vogel entanglement
criterion \cite{vogel's ent criterion}. In Ref. \cite{Adam-criterion,Adam-generalizes voge;},
Miranowicz et al. have explicitly established this point. 

So far we have described criteria of nonclassicality that are related
to the lowest order nonclassicality. However, nonclassicality may
be witnessed via higher order criterion of nonclassicality, too. Investigations
on higher order nonclassical properties of various optical systems
have been performed since long. For example, in late seventies some
of the present authors showed the existence of higher order nonclassical
photon statistics in different optical systems using criterion based
on higher order moments of number operators (cf. Ref. \cite{perina-review}
and Chapter 10 of \cite{Perina-Book} and references therein). However,
in those early works, higher order antibunching (HOA) was not specifically
discussed, but existence of higher order nonclassical photon statistics
was reported for degenerate and nondegenerate parametric processes
in single and compound signal-idler modes, respectively and also for
Raman scattering in compound Stokes-anti-Stokes mode up to $n=5$.
Criterion for HOA was categorically introduced by C. T. Lee \cite{C T Lee}
in 1990. Since then HOA is reported in several quantum optical systems
(\cite{HOAis not rare,generalized-higher order} and references therein)
and atomic systems \cite{pathak-pra2}. However, except a recent effort
by us \cite{kishore-1} no effort has yet been made to study HOA in
optical couplers. The existence of HOA can be witnessed through a
set of equivalent but different criteria, all of which can be interpreted
as modified Lee criterion. In what follows we will investigate the
existence of HOA in the contradirectional optical coupler of our interest
using a simple criterion of $n$-th order single-mode antibunching
introduced by some of us \cite{HOAwithMartin} \begin{equation}
\begin{array}{lcl}
D_{a}(n)=\left\langle a^{\dagger n}a^{n}\right\rangle -\left\langle a^{\dagger}a\right\rangle ^{n} & < & 0.\end{array}\label{hoa-1}\end{equation}
Here $n=2$ and $n\geq3$ refer to the usual antibunching and the
higher order antibunching, respectively. 

Similarly, we can also investigate the existence of higher order squeezing,
but definition of higher order squeezing is not unique. To be precise,
it is usually investigated using two different criteria \cite{HIlery-amp-sq,Hong-Mandel1,HOng-mandel2}.
The first criterion was introduced by Hong and Mandel in 1985 \cite{Hong-Mandel1,HOng-mandel2}
and a second criterion was subsequently introduced by Hillery in 1987
\cite{HIlery-amp-sq}. In Hong and Mandel criterion \cite{Hong-Mandel1,HOng-mandel2},
the reduction of higher order moments of usual quadrature operators
with respect to their coherent state counterparts are considered as
higher order squeezing. However, in Hillery's criterion, reduction
of variance of an amplitude powered quadrature variable for a quantum
state with respect to its coherent state counterpart is considered
as higher order squeezing. In what follows we have restricted our
study on higher order squeezing to Hillery's criterion of amplitude
powered squeezing. Specifically, Hillery introduced amplitude powered
quadrature variables as \begin{equation}
Y_{1,a}=\frac{a^{n}+\left(a^{\dagger}\right)^{n}}{2}\label{eq:quadrature-power1}\end{equation}
 and \begin{equation}
Y_{2,a}=i\left(\frac{\left(a^{\dagger}\right)^{n}-a^{n}}{2}\right).\label{eq:quadrature-power2}\end{equation}
It is easy to check that $Y_{1,a}$ and $Y_{2,a}$ do not commute
and consequently we can obtain an uncertainty relation and thus a
criterion of $n^{th}$ order amplitude squeezing as\begin{equation}
{\normalcolor {\color{green}{\normalcolor A_{i,a}=\left(\Delta Y_{i,a}\right)^{2}-}{\normalcolor \frac{1}{2}}{\normalcolor \left|\left\langle \left[Y_{1,a},Y_{2,a}\right]\right\rangle \right|}{\normalcolor <0}}.}\end{equation}
Thus, for the specific case, $n=2$, Hillery's criterion for amplitude
squared squeezing can be obtained as \begin{equation}
A_{i,a}=\left(\Delta Y_{i,a}\right)^{2}-\left\langle N_{a}+\frac{1}{2}\right\rangle <0,\label{eq:criterion-amplitude squared}\end{equation}
where $i\in\{1,2\}.$ Similarly, we can obtain specific criteria of
amplitude powered squeezing for other values of $n.$ Further, there
exists a set of higher order inseparability criteria. To be precise,
all the criteria that are used for witnessing the existence of multi-partite
(multi-mode) entanglement are essentially higher order criteria \cite{higher order multiparty1,higher order multyparty2,higher order multyparty3}
as they always uncover some higher order correlation. Interestingly,
even in bipartite (two-mode) case one can introduce operational criterion
for detection of higher order entanglement. For example, Hillery-Zubairy
introduced following two criteria of higher order intermodal entanglement
\cite{HZ-PRL} \begin{equation}
E_{ab}^{m,n}=\left\langle \left(a^{\dagger}\right)^{m}a^{m}\left(b^{\dagger}\right)^{n}b^{n}\right\rangle -\left\vert \left\langle a^{m}\left(b^{\dagger}\right)^{n}\right\rangle \right\vert ^{2}<0,\label{hoe-criteria-1}\end{equation}
and\begin{equation}
E_{ab}^{\prime m,n}=\left\langle \left(a^{\dagger}\right)^{m}a^{m}\right\rangle \left\langle \left(b^{\dagger}\right)^{n}b^{n}\right\rangle -\left\vert \left\langle a^{m}b^{n}\right\rangle \right\vert ^{2}<0.\label{hoe-criteria-2}\end{equation}
Here $m$ and $n$ are non-zero positive integers and the lowest possible
values of $m$ and $n$ are $m=n=1$ which reduces (\ref{hoe-criteria-1})
and (\ref{hoe-criteria-2}) to usual HZ-I criterion (i.e., (\ref{hz1}))
and HZ-II criterion (i.e., (\ref{hz2})), respectively. Thus these
two criteria may be viewed as generalized versions of the well known
lower order criteria of Hillery and Zubairy (i.e., the criteria described
in (\ref{hz1}) and (\ref{hz2})). However, these generalized criterion
can also be obtained as special cases of more general criterion of
Shchukin and Vogel \cite{vogel's ent criterion}. For the convenience
of the readers, we refer to (\ref{hoe-criteria-1}) and (\ref{hoe-criteria-2})
as HZ-I criterion and HZ-II criterion respectively in analogy to the
lowest order cases. In what follows, a quantum state will be called
(bipartite) higher order entangled state if it is found to satisfy
(\ref{hoe-criteria-1}) and/or (\ref{hoe-criteria-2}) for any choice
of integers $m$ and $n$ satisfying $m+n\geq3.$ Existence of higher
order entanglement can also be viewed through the criteria of multi-partite
entanglement. For example, Li et al. \cite{Ent condition-multimode}
proved that a three-mode (tripartite) quantum state is not bi-separable
in the form $ab_{1}|b_{2}$ (i.e., compound mode $ab_{1}$ is entangled
with the mode $b_{2}$) if the following inequality holds for the
three-mode system 

\begin{equation}
E_{ab_{1}|b_{2}}^{m,n,l}=\langle\left(a^{\dagger}\right)^{m}a^{m}\left(b_{1}^{\dagger}\right)^{n}b_{1}^{n}\left(b_{2}^{\dagger}\right)^{l}b_{2}^{l}\rangle-|\langle a^{m}b_{1}^{n}(b_{2}^{\dagger})^{l}\rangle|^{2}<0,\label{eq:tripartite ent1}\end{equation}
where $m,\, n,\, l$ are positive integers and annihilation operators
$a,b_{1},b_{2}$ correspond to the three modes. A quantum state that
satisfy  (\ref{eq:tripartite ent1}) is referred to as $ab_{1}|b_{2}$
entangled state. The three-mode inseparability criterion mentioned
above is not unique. There exist various alternative criteria of three-mode
entanglement. For example, an alternative criterion for detection
of $ab_{1}|b_{2}$ entangled state is \cite{Ent condition-multimode}
\begin{equation}
E_{ab_{1}|b_{2}}^{\prime m,n,l}=\langle\left(a^{\dagger}\right)^{m}a^{m}\left(b_{1}^{\dagger}\right)^{n}b_{1}^{n}\rangle\langle\left(b_{2}^{\dagger}\right)^{l}b_{2}^{l}\rangle-|\langle a^{m}b_{1}^{n}b_{2}^{l}\rangle|^{2}<0.\label{eq:tripartite ent2}\end{equation}
For $m=n=l=1,$ this criterion coincides with Miranowicz et al.'s
criterion \cite{Adam-generalizes voge;}. Using (\ref{eq:tripartite ent1})
and (\ref{eq:tripartite ent2}), we can easily obtain criteria for
detection of $a|b_{1}b_{2}$ and $b_{1}|ab_{2}$ entangled states
and use them to obtain a simple criterion for detection of fully entangled
tripartite state. Specifically, using (\ref{eq:tripartite ent1})
and (\ref{eq:tripartite ent2}) respectively we can conclude that
the three modes of our interest are not bi-separable in any form if
one of the following two sets of inequalities are satisfied simultaneously
\begin{equation}
E_{ab_{1}|b_{2}}^{1,1,1}<0,\, E_{a|b_{1}b_{2}}^{1,1,1}<0,\, E_{b_{1}|b_{2}a}^{1,1,1}<0,\label{eq:fully enta 0}\end{equation}

\begin{equation}
E_{ab_{1}|b_{2}}^{\prime1,1,1}<0,\, E_{a|b_{1}b_{2}}^{\prime1,1,1}<0,\, E_{b_{1}|b_{2}a}^{\prime1,1,1}<0.\label{eq:fully enta 1}\end{equation}
 Further, for a fully separable pure state we always have \begin{equation}
|\langle ab_{1}b_{2}\rangle|=|\langle a\rangle\langle b_{1}\rangle\langle b_{2}\rangle|\leq\left[\langle N_{a}\rangle\langle N_{b_{1}}\rangle\langle N_{b_{2}}\rangle\right]^{\frac{1}{2}}.\label{eq:fully ent2}\end{equation}
Thus, a three-mode pure state that violates (\ref{eq:fully ent2})
(i.e., satisfies $\langle N_{a}\rangle\langle N_{b_{1}}\rangle\langle N_{b_{2}}\rangle-|\langle ab_{1}b_{2}\rangle|^{2}<0)$
and simultaneously satisfies either (\ref{eq:fully enta 0}) or (\ref{eq:fully enta 1})
is a fully entangled state as it is neither fully separable nor bi-separable
in any form.

\section{{\normalsize Nonclassicality in contradirectional optical coupler\label{sec:Nonclassicality-in-contradirectional}}}

Spatial evolution of different operators that are relevant for witnessing
nonclassicality can be obtained using the perturbative solutions (\ref{2})-(\ref{eq:soln})
reported here. For example, using (\ref{2})-(\ref{eq:soln}) we can
obtain the following closed form expressions for number operators
of various field modes \begin{equation}
\begin{array}{lcl}
N_{a} & = & a^{\dagger}a=|f_{1}|^{2}a^{\dagger}(L)a(L)+|f_{2}|^{2}b_{1}^{\dagger}(0)b_{1}(0)+\left[f_{1}^{*}f_{2}a^{\dagger}(L)b_{1}(0)+f_{1}^{*}f_{3}a^{\dagger}(L)b_{1}^{\dagger}(0)b_{2}(0)\right.\\
 & + & \left.f_{1}^{*}f_{4}a^{\dagger2}(L)b_{2}(0)+f_{2}^{*}f_{3}b_{1}^{\dagger2}(0)b_{2}(0)+f_{2}^{*}f_{4}b_{1}^{\dagger}(0)a^{\dagger}(L)b_{2}(0)+{\rm h.c.}\right],\end{array}\label{eq:na}\end{equation}
 \begin{equation}
\begin{array}{lcl}
N_{b_{1}} & = & b_{1}^{\dagger}b_{1}=|g_{1}|^{2}a^{\dagger}(L)a(L)+|g_{2}|^{2}b_{1}^{\dagger}(0)b_{1}(0)+\left[g_{1}^{*}g_{2}a^{\dagger}(L)b_{1}(0)+g_{1}^{*}g_{3}a^{\dagger}(L)b_{1}^{\dagger}(0)b_{2}(0)\right.\\
 & + & \left.g_{1}^{*}g_{4}a^{\dagger2}(L)b_{2}(0)+g_{2}^{*}g_{3}b_{1}^{\dagger2}(0)b_{2}(0)+g_{2}^{*}g_{4}b_{1}^{\dagger}(0)a^{\dagger}(L)b_{2}(0)+{\rm h.c.}\right],\end{array}\label{eq:nb}\end{equation}
 \begin{equation}
N_{b_{2}}=b_{2}^{\dagger}b_{2}=b_{2}^{\dagger}(0)b_{2}(0)+\left[h_{2}b_{2}^{\dagger}(0)b_{1}^{2}(0)+h_{3}b_{2}^{\dagger}(0)b_{1}(0)a(L)+h_{4}b_{2}^{\dagger}(0)a^{2}(L)+{\rm h.c.}\right].\label{eq:nb2}\end{equation}
It is now straight forward to compute the average values of the number
of photons in different modes with respect to a given initial state.
In the present work we consider that the initial state is product
of three coherent states: $|\alpha\rangle|\beta\rangle|\gamma\rangle,$
where $|\alpha\rangle,\,|\beta\rangle$ and $|\gamma\rangle$ are
eigen kets of annihilation operators $a,\, b_{1}$ and $b_{2}$, respectively.
Thus, \begin{equation}
a(L)|\alpha\rangle|\beta\rangle|\gamma\rangle=\alpha|\alpha\rangle|\beta\rangle|\gamma\rangle,\label{2bcd}\end{equation}
and $|\alpha|^{2},\,|\beta|^{2},\,|\gamma|^{2}$ are the number of
input photons in the field modes $a,\, b_{1}$ and $b_{2}$, respectively.
For a spontaneous process, $\beta=\gamma=0$ and $\alpha\ne0.$ Whereas,
for a stimulated process, the complex amplitudes are not necessarily
zero and it seems reasonable to consider $\alpha>\beta>\gamma.$

\subsection{{\normalsize Single-mode and intermodal squeezing}}

\begin{figure}
\begin{centering}
\includegraphics[angle=-90,scale=0.5]{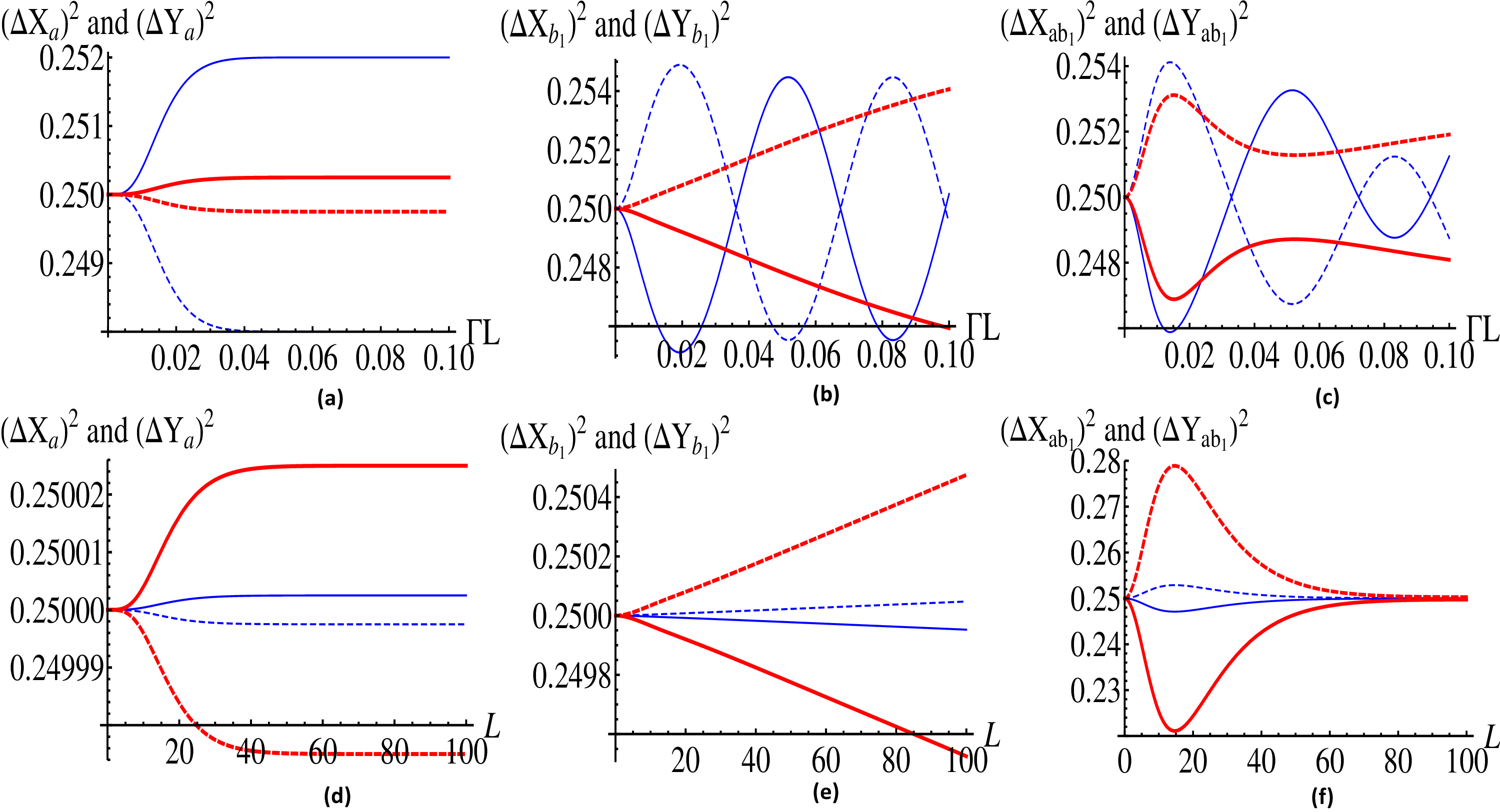}
\par\end{centering}

\caption{\label{fig:Squeezing-del k-G}(color online) Existence of quadrature
squeezing in modes $a$ and $b_{1}$ and intermodal squeezing in mode
$ab_{1}$ is illustrated with $k=0.1,\,\alpha=5,\beta=2,\gamma=1$
for various values of phase mismatching $\Delta k$ and nonlinear
coupling constant $\Gamma$. In Fig. (a)-(c) squeezing and intermodal
squeezing is plotted with rescaled interaction length $\Gamma L$
with $\Gamma=0.001$ for $\Delta k=10^{-1}$ (thin blue lines) and
$\Delta k=10^{-2}$ (thick red lines). In Fig. (d)-(f) squeezing and
intermodal squeezing is plotted with interaction length $L$ with
$\Delta k=10^{-4}$ for $\Gamma=0.001$ (thin blue lines) and $\Gamma=0.01$
(thick red lines). In all the sub-figures a solid (dashed) line represents
$X_{i}$ $(Y_{i})$ where $i\in\left\{ a,b_{1}\right\} $ or $X_{ab_{1}}$
$(Y_{ab_{1}})$ quadrature. Parts of the plots that depict values
of variance <$\frac{1}{4}$ in (a) and (d) show squeezing in quadrature
variable $Y_{a}$, that in (b) and (c) show squeezing in quadrature
variable $X_{b_{1}}$, $Y_{b_{1}}$ and intermodal squeezing in quadrature
variable $X_{ab_{1}}$, $Y_{ab_{1}},$ respectively. Similarly, (e)
and (f) show squeezing and intermodal squeezing in quadrature variable
$X_{b_{1}}$ and $X_{ab_{1}}$ respectively. Squeezing in the other
quadrature variables (say $X_{a}$) can be obtained by suitable choice
of phases of the input coherent states.}
\end{figure}

Using (\ref{2})-(\ref{eq:soln}) and (\ref{eq:condition for squeezing})-(\ref{eq:two-mode quadrature})
we obtain analytic expressions for variance in single-mode and compound
mode quadratures as

\begin{equation}
\begin{array}{lcl}
\left[\begin{array}{c}
\left(\Delta X_{a}\right)^{2}\\
\left(\Delta Y_{a}\right)^{2}\end{array}\right] & = & \frac{1}{4}\left[1\pm\left\{ \left(f_{1}f_{4}+f_{2}f_{3}\right)\gamma+{\rm c.c}.\right\} \right],\\
\left[\begin{array}{c}
\left(\Delta X_{b_{1}}\right)^{2}\\
\left(\Delta Y_{b_{1}}\right)^{2}\end{array}\right] & = & \frac{1}{4}\left[1\pm\left\{ \left(g_{1}g_{4}+g_{2}g_{3}\right)\gamma+{\rm c.c}.\right\} \right],\\
\left[\begin{array}{c}
\left(\Delta X_{b_{2}}\right)^{2}\\
\left(\Delta Y_{b_{2}}\right)^{2}\end{array}\right] & = & \frac{1}{4},\end{array}\label{eq:squeezing}\end{equation}
 and 

\begin{equation}
\begin{array}{lcl}
\left[\begin{array}{c}
\left(\Delta X_{ab_{1}}\right)^{2}\\
\left(\Delta Y_{ab_{1}}\right)^{2}\end{array}\right] & = & \frac{1}{4}\left[1\pm\frac{1}{2}\left\{ \left(\left(f_{1}+g_{1}\right)\left(f_{4}+g_{4}\right)+\left(f_{2}+g_{2}\right)\left(f_{3}+g_{3}\right)\right)\gamma+{\rm c.c}.\right\} \right],\\
\left[\begin{array}{c}
\left(\Delta X_{ab_{2}}\right)^{2}\\
\left(\Delta Y_{ab_{2}}\right)^{2}\end{array}\right] & = & \frac{1}{4}\left[1\pm\frac{1}{2}\left\{ \left(f_{1}f_{4}+f_{2}f_{3}\right)\gamma+{\rm c.c}.\right\} \right]=\frac{1}{2}\left[\begin{array}{c}
\left(\Delta X_{a}\right)^{2}\\
\left(\Delta Y_{a}\right)^{2}\end{array}\right]+\frac{1}{8},\\
\left[\begin{array}{c}
\left(\Delta X_{b_{1}b_{2}}\right)^{2}\\
\left(\Delta Y_{b_{1}b_{2}}\right)^{2}\end{array}\right] & = & \frac{1}{4}\left[1\pm\frac{1}{2}\left\{ \left(g_{1}g_{4}+g_{2}g_{3}\right)\gamma+{\rm c.c}.\right\} \right]=\frac{1}{2}\left[\begin{array}{c}
\left(\Delta X_{b_{1}}\right)^{2}\\
\left(\Delta Y_{b_{1}}\right)^{2}\end{array}\right]+\frac{1}{8},\end{array}\label{eq:intermodal squeezing}\end{equation}
respectively. From Eq. (\ref{eq:squeezing}) it is clear that no squeezing
is observed in $b_{2}$ mode. However, squeezing is possible in $a$
mode and $b_{1}$ mode as illustrated in Fig. \ref{fig:Squeezing-del k-G}
a-b, d-e. Further, we observed intermodal squeezing in quadratures
$X_{ab_{1}}$ and $Y_{ab_{1}}$ by plotting right hand sides of Eq.
(\ref{eq:intermodal squeezing}) in Fig. \ref{fig:Squeezing-del k-G}
c and Fig. \ref{fig:Squeezing-del k-G} f. Variation of amount of
squeezing in different modes with phase mismatch $\Delta k$ and nonlinear
coupling constant $\Gamma$ are shown in Fig. \ref{fig:Squeezing-del k-G}
a-c and Fig. \ref{fig:Squeezing-del k-G} d-f, respectively. We have
also studied the effect of linear coupling constant $k$ on the amount
of squeezing, but its effect is negligible in all other quadratures
except the quadratures of compound mode $ab_{1}.$ In compound mode
$ab_{1}$, after a short distance depth of squeezing is observed to
increase with the decrease in the linear coupling constant $k$ (this
is not illustrated through figure). Intermodal squeezing in compound
mode quadratures $Y_{ab_{2}}$ and $X_{b_{1}b_{2}}$ can be visualized
from the last two rows of Eq.(\ref{eq:intermodal squeezing}). Specifically,
 we can see that variance in compound mode quadrature $X_{jb_{2}}$
and $Y_{jb_{2}}$ have \emph{bijective }(both one-to-one and onto)
correspondence with the variance in $X_{j}$ and $Y_{j}$, respectively,
where $j\in\left\{ a,b_{1}\right\} .$ To be precise, quadrature squeezing
in single-mode $X_{j}$($Y_{j}$) implies quadrature squeezing in
$X_{j,b_{2}}(Y_{j,b_{2}})$ and vice versa. For example, $\left(\Delta X_{jb_{2}}\right)^{2}<\frac{1}{4}\Rightarrow\frac{1}{2}\left(\Delta X_{j}\right)^{2}+\frac{1}{8}<\frac{1}{4}$
or, $\left(\Delta X_{j}\right)^{2}<2\left(\frac{1}{4}-\frac{1}{8}\right)=\frac{1}{4}.$
This is why we have not explicitly shown the variance of compound
mode quadratures $X_{j,b_{2}}$ and $Y_{j,b_{2}}$ with different
parameters as we have done for the other cases. As we have shown squeezing
in $Y_{a},\, X_{b_{1}}$ and $Y_{b_{1}}$ through Fig. \ref{fig:Squeezing-del k-G}
a-b, and c-d this implies the existence of squeezing in quadrature
$Y_{ab_{2}},\, X_{b_{1}b_{2}}$ and $Y_{b_{1}b_{2}}.$ Thus we have
observed intermodal squeezing in compound modes involving $b_{2}$.
This nonclassical feature was not observed in earlier studies \cite{perina-review}
as in those studies $b_{2}$ mode was considered classical. The plots
do not show quadrature squeezing in $X_{a},$ $X_{ab_{2}}$, and $X_{ab_{1}}$
(for some specific values of $\Gamma$). However, a suitable choice
of phase of the input coherent state would lead to squeezing in these
quadratures. For example, if we replace $\gamma$ by $-\gamma$ (i.e.,
if we chose $\gamma=\exp(i\pi)$ instead of present choice of $\gamma=1$)
then we would observe squeezing in all these quadratures, but the
squeezing that is observed now with the original choice of $\gamma$
would vanish. This is so as all the expressions of variance of quadrature
variables that are $\neq\frac{1}{4}$ have a common functional form:
$\frac{1}{4}\pm\gamma F(f_{i},g_{i})$ (c.f., Eqs. (\ref{eq:squeezing})
and (\ref{eq:intermodal squeezing}) ). 

\subsection{{\normalsize Higher order squeezing}}

\begin{figure}
\begin{centering}
\subfloat[]{

\includegraphics[scale=0.5]{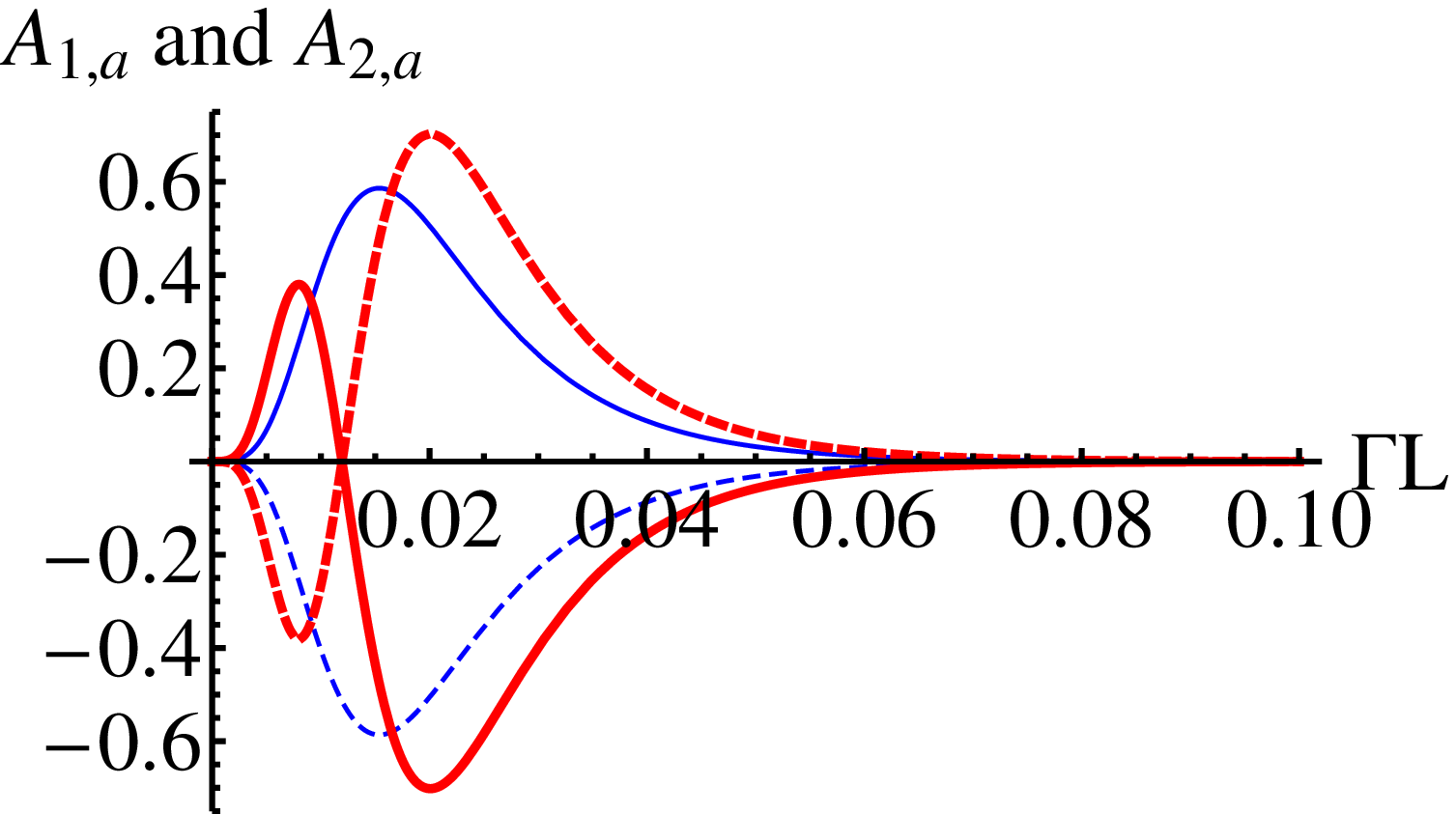}}\subfloat[]{

\includegraphics[scale=0.5]{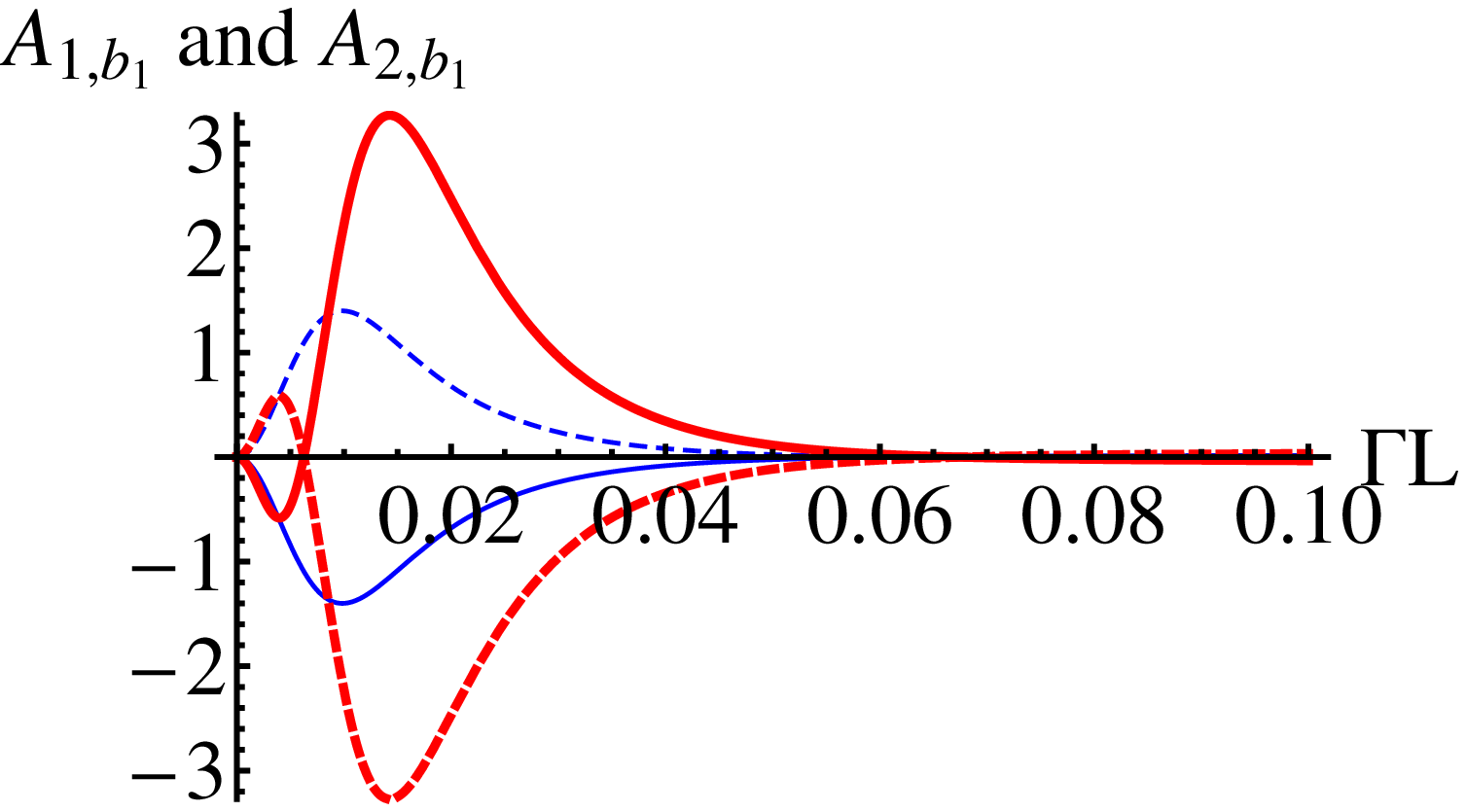}

}
\par\end{centering}

\centering{}\caption{{\large \label{fig:Amplitude-squared-squeezing}}(color online) Amplitude
powered squeezing is observed in (a) $a$ mode and (b) $b_{1}$ mode
for $k=0.1,\,\Gamma=0.001,\,\Delta k=10^{-4},\,\alpha=3,\beta=2,\gamma=1.$
Negative parts of the solid line represent amplitude powered squeezing
in quadrature variable $Y_{1,a}$ ($Y_{1,b_{1}})$ and that in the
dashed line represents squeezing in quadrature variable $Y_{2,a}$
($Y_{2,b_{1}})$ for $n=2$ (thin blue lines) and $n=3$ (thick red
lines). To display the plots in the same scale $A_{i,a}$ and $A_{i,b_{1}}$
for $n=2$ are multiplied by 10, where $i\in\left\{ 1,2\right\} $.}
\end{figure}

After establishing the existence of squeezing in single-modes and
compound modes we now examine the possibility of higher order squeezing
using Eqs. (\ref{2})-(\ref{eq:soln}), (\ref{eq:na})-(\ref{eq:nb2})
and (\ref{eq:criterion-amplitude squared}). In fact, we obtain \begin{equation}
\begin{array}{lcl}
\left[\begin{array}{c}
A_{1,a}\\
A_{2,a}\end{array}\right] & = & \pm\frac{n^{2}}{4}\left[\gamma\left(f_{1}f_{4}+f_{2}f_{3}\right)\left(f_{1}\alpha+f_{2}\beta\right)^{2n-2}+{\rm c.c.}\right],\end{array}\label{eq:asq-1}\end{equation}

\begin{equation}
\begin{array}{lcl}
\left[\begin{array}{c}
A_{1,b_{1}}\\
A_{2,b_{1}}\end{array}\right] & = & \pm\frac{n^{2}}{4}\left[\gamma\left(g_{1}g_{4}+g_{2}g_{3}\right)\left(g_{1}\alpha+g_{2}\beta\right)^{2n-2}+{\rm c.c.}\right],\end{array}\label{eq:as-q-2}\end{equation}
 and \begin{equation}
\begin{array}{lcl}
\left[\begin{array}{c}
A_{1,b_{2}}\\
A_{2,b_{2}}\end{array}\right] & = & 0.\end{array}\label{eq:as-q-3}\end{equation}
Thus, we do not get any signature of amplitude powered squeezing in
$b_{2}$ mode using the present solution. In contrary, mode $a$ $(b_{1})$
is found to show amplitude powered squeezing in one of the quadrature
variables for any value of interaction length as both $A_{1,a}$ and
$A_{2,a}$ ($A_{1,b_{1}}$ and $A_{2,b_{1}}$) cannot be positive
simultaneously. To study the possibilities of amplitude powered squeezing
in further detail we have plotted the spatial variation of $A_{i,a}$
and $A_{i,b_{1}}$ in Fig. \ref{fig:Amplitude-squared-squeezing}.
Negative regions of these two plots clearly illustrate the existence
of amplitude powered squeezing in both $a$ and $b_{1}$ modes for
$n=2$ and $n=3$. Extending our observations in context of single-mode
squeezing and intermodal squeezing we can state that the appearance
of amplitude powered squeezing in a particular quadrature can be controlled
by suitable choice of phase of input coherent state $\gamma$ as the
expressions for amplitude powered squeezing reported in (\ref{eq:asq-1})
and (\ref{eq:as-q-2}) have a common functional form $\pm\gamma F(f_{i},g_{i})$. 

\subsection{{\normalsize Lower order and higher order antibunching}}

\begin{figure}
\centering{}\includegraphics[angle=-90, scale=0.5]{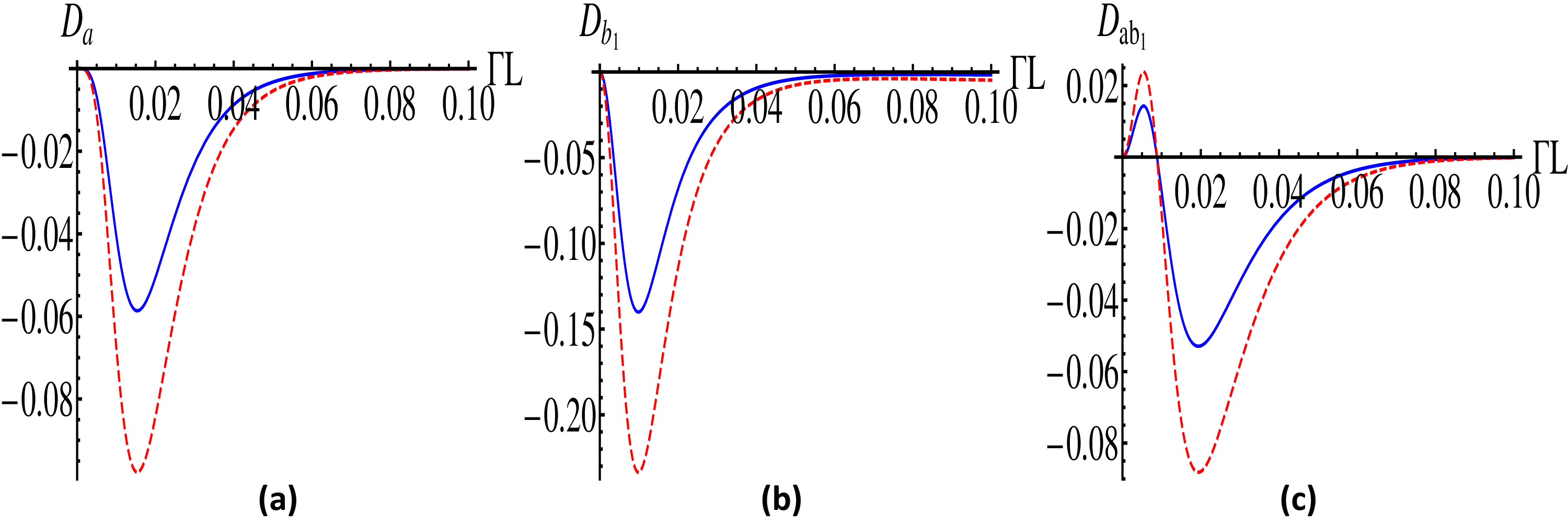}\caption{\label{fig:Antibunching} (color online) Variation of $D_{i}$ and
$D_{ij}$ with rescaled interaction length $\lyxmathsym{\textgreek{G}}L$
for $\alpha=3$ (smooth line) and $\alpha=5$ (dashed line) in (a)
single mode $a$, (b) single mode $b_{1}$ and (c) compound mode $ab_{1}$
with $k=0.1,\,\Gamma=0.001,\,\Delta k=10^{-4},\,\beta=2$ and $\gamma=1$
for (a) and $\gamma=-1$ for (b)-(c). Negative parts of the plots
illustrate the existence of nonclassical photon statistics (antibunching).}
\end{figure}

The condition of HOA is already provided through the inequality (\ref{hoa-1}).
Now using this inequality along with Eqns. (\ref{2})-(\ref{eq:soln})
and (\ref{eq:na})-(\ref{eq:nb2}) we can obtain closed form analytic
expressions for $D_{i}(n)$ for various modes as follows \begin{equation}
\begin{array}{lcl}
D_{a}(n) & = & ^{n}C_{2}\gamma|\left(f_{1}\alpha+f_{2}\beta\right)|^{2n-4}\left\{ \left(f_{1}\alpha+f_{2}\beta\right)^{2}\left(f_{2}^{*}f_{3}^{*}+f_{1}^{*}f_{4}^{*}\right)+{\rm c.c.}\right\} ,\end{array}\label{eq:dan}\end{equation}
\begin{equation}
\begin{array}{lcl}
D_{b_{1}}(n) & = & ^{n}C_{2}\gamma|\left(g_{1}\alpha+g_{2}\beta\right)|^{2n-4}\left\{ \left(g_{1}\alpha+g_{2}\beta\right)^{2}\left(g_{2}^{*}g_{3}^{*}+g_{1}^{*}g_{4}^{*}\right)+{\rm c.c.}\right\} ,\end{array}\label{eq:dbn}\end{equation}

\begin{equation}
D_{b_{2}}(n)=0.\label{eq:db2n}\end{equation}
Further, using the condition of intermodal antibunching described
in (\ref{eq:antib2}) and Eqns. (\ref{2})-(\ref{eq:soln}) we obtain
following closed form expressions of $D_{ij}$ \begin{equation}
\begin{array}{lcl}
D_{ab_{1}} & = & \left\{ \left(|g_{1}|^{2}f_{1}^{*}f_{4}+f_{1}^{*}f_{3}g_{1}^{*}g_{2}\right)\alpha^{*2}\gamma+\left(|g_{2}|^{2}f_{2}^{*}f_{3}+f_{2}^{*}f_{4}g_{2}^{*}g_{1}\right)\beta^{*2}\gamma+\left(|g_{1}|^{2}-|g_{2}|^{2}\right)\left(f_{2}^{*}f_{4}-f_{1}^{*}f_{3}\right)\alpha\beta\gamma^{*}+{\rm c.c.}\right\} ,\end{array}\label{eq:dab1}\end{equation}
\begin{equation}
\begin{array}{lcl}
D_{ab_{2}} & = & 0,\end{array}\label{eq:dab2}\end{equation}

\begin{equation}
D_{b_{1}b_{2}}=0.\label{eq:db1b2}\end{equation}
From the above expressions it is clear that neither the single-mode
antibunching nor the intermodal antibunching is obtained involving
$b_{2}$ mode. As the expressions obtained in the right hand sides
of (\ref{eq:dan}), (\ref{eq:dbn}) and (\ref{eq:dab1}) are not simple,
we plot them to investigate the existence of single-mode and compound
mode antibunching. The plots for usual antibunching and intermodal
antibunching are shown in Fig. \ref{fig:Antibunching}. Existence
of antibunching is obtained in single-mode $a$ for $\gamma=1$ and
the same is illustrated through Fig. \ref{fig:Antibunching} a. However,
in a effort to obtain antibunching in $b_{1}$ mode, we do not observe
any antibunching in $b_{1}$ mode for $\gamma=1$. Interestingly,
from (\ref{eq:dbn}) it is clear that if we replace $\gamma=1$ by
$\gamma=\exp(i\pi)=-1$ as before and keep $\alpha,\beta$ unchanged,
then we would observe antibunching for all values of rescaled interaction
length $\Gamma L.$ This is true in general for all values of $\gamma$.
To be precise, if we observe antibunching (bunching) in a mode for
$\gamma=c$ we will always observe bunching (antibunching) for $\gamma=-c,$
if we keep $\alpha,\beta$ unchanged. This fact is illustrated through
Fig. \ref{fig:Antibunching} b where we plot variation of $D_{b_{1}}$
with $\Gamma L$ and have observed the existence of antibunching.
In compound mode $ab_{1},$ we can observe existence of antibunching
for both $\gamma=1$ and $\gamma=-1$. However, in Fig. \ref{fig:Antibunching}
c we have illustrated the existence of intermodal antibunching in
compound mode $ab_{1}$ by plotting variation $D_{ab_{1}}$ with rescaled
interaction length $\Gamma L$ for $\gamma=-1$ as region of nonclassicality
is relatively larger (compared to the case where $\gamma=1$ and $\alpha,\beta$
are same) in this case.

\begin{figure}
\begin{centering}
\includegraphics[angle=-90,scale=0.5]{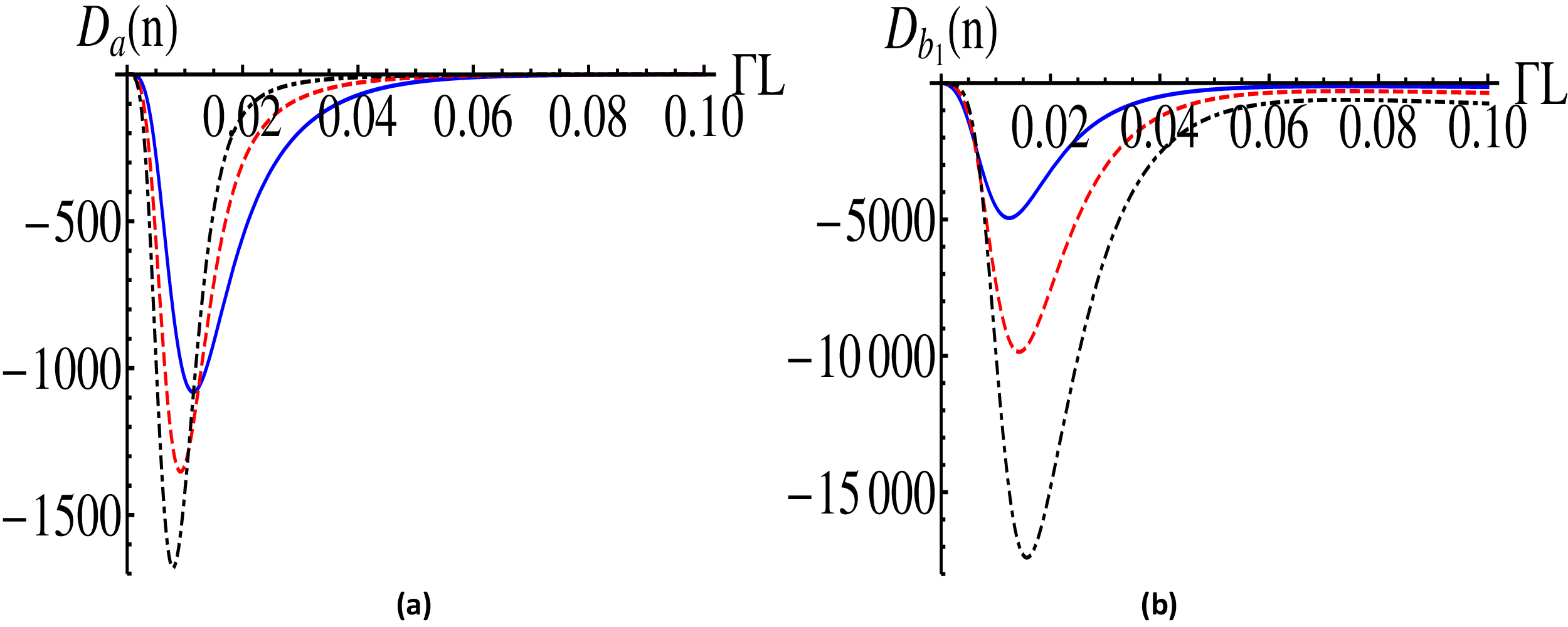}
\par\end{centering}

\centering{}\caption{\label{fig:HOA} (color online) Variation of $D_{i}(n)$ with rescaled
interaction length $\Gamma L$ in (a) mode $a$ with $k=0.1,\,\Gamma=0.001,\,\Delta k=10^{-4},\,\alpha=5,\,\beta=2,\,\gamma=1$
and (b) mode $b_{1}$ with $k=0.1,\,\Gamma=0.001,\,\Delta k=10^{-4},\,\alpha=5,\,\beta=2,\,\gamma=-1$
for $n=3$ (smooth lines), $n=4$ (dashed lines) and $n=5$ (dot-dashed
lines). To display the plots in the same scale $D_{i}(3)$ and $D_{i}(4)$
are multiplied by 400 and 20 respectively, where $i\in\left\{ a,b_{1}\right\} $.\textcolor{red}{{}
}Negative parts of the plots show HOA.}
\end{figure}

Now we may extend the discussion to HOA and plot right hand sides
of (\ref{eq:dan}) and (\ref{eq:dbn}) for various values of $n$.
The plots are shown in Fig. \ref{fig:HOA} which clearly illustrates
the existence of HOA and also demonstrate that the depth of nonclassicality
increases with $n.$ This is consistent with earlier observations
on HOA in other systems \cite{HOAwithMartin}.

\subsection{{\normalsize Lower order and higher order intermodal entanglement}}

\begin{figure}
\begin{centering}
\includegraphics[scale=0.5]{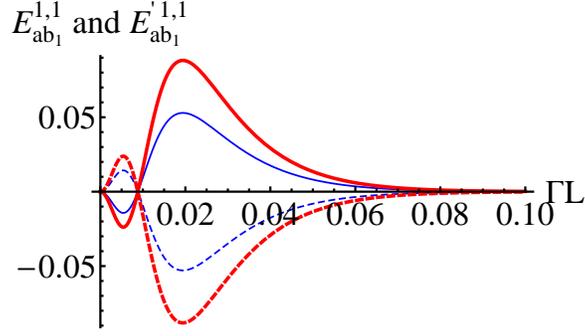}
\par\end{centering}

\caption{{\large \label{fig:ent-hz1-II}}(color online) Hillery-Zubairy criterion
I (solid line) and criterion II (dashed line) for entanglement are
showing intermodal entanglement between modes $a$ and $b_{1}.$ Here
$E_{ab_{1}}^{1,1}$ (solid line) and $E_{ab_{1}}^{\prime1,1}$ (dashed
line) are plotted with rescaled interaction length $\lyxmathsym{\textgreek{G}}L$
for mode $ab_{1}$ with $k=0.1,\,\Gamma=0.001,\,\Delta k=10^{-4},\,\beta=2,\gamma=1$
for $\alpha=3$ (thin blue lines) and $\alpha=5$ (thick red lines).}
\end{figure}

We first examine the existence of intermodal entanglement in compound
mode $ab_{1}$ using HZ-I criterion (\ref{hz1}). To do so we use
Eqns. (\ref{2})-(\ref{eq:soln}) and (\ref{eq:na})-(\ref{eq:nb2})
and obtain\begin{equation}
\begin{array}{lcl}
E_{ab_{1}}^{1,1} & = & \langle N_{a}N_{b_{1}}\rangle-|\langle ab_{1}^{\dagger}\rangle|^{2}\\
 & = & \left(|g_{1}|^{2}f_{4}^{*}f_{1}+f_{3}^{*}f_{1}g_{2}^{*}g_{1}\right)\alpha^{2}\gamma^{*}+\left(|f_{1}|^{2}g_{1}^{*}g_{4}+f_{1}^{*}f_{2}g_{1}^{*}g_{3}\right)\alpha^{*2}\gamma\\
 & + & \left(|g_{2}|^{2}f_{3}^{*}f_{2}+f_{4}^{*}f_{2}g_{1}^{*}g_{2}\right)\beta^{2}\gamma^{*}+\left(|f_{2}|^{2}g_{2}^{*}g_{3}+f_{2}^{*}f_{1}g_{2}^{*}g_{4}\right)\beta^{*2}\gamma\\
 & + & \left(|g_{1}|^{2}-|g_{2}|^{2}\right)\left\{ \left(f_{4}^{*}f_{2}-f_{3}^{*}f_{1}\right)\alpha\beta\gamma^{*}-\left(g_{2}^{*}g_{4}-g_{1}^{*}g_{3}\right)\alpha^{*}\beta^{*}\gamma\right\} .\end{array}\label{eq:HZ-1-ab1}\end{equation}
Similarly, using HZ-II criterion (\ref{hz2}) we obtain

\begin{equation}
\begin{array}{lcl}
E_{ab_{1}}^{\prime1,1} & = & \langle N_{a}\rangle\langle N_{b_{1}}\rangle-|\langle ab_{1}\rangle|^{2}\\
 & = & -\left[\left(|g_{1}|^{2}f_{4}^{*}f_{1}+f_{3}^{*}f_{1}g_{2}^{*}g_{1}\right)\alpha^{2}\gamma^{*}+\left(|f_{1}|^{2}g_{1}^{*}g_{4}+f_{1}^{*}f_{2}g_{1}^{*}g_{3}\right)\alpha^{*2}\gamma\right.\\
 & + & \left.\left(|g_{2}|^{2}f_{3}^{*}f_{2}+f_{4}^{*}f_{2}g_{1}^{*}g_{2}\right)\beta^{2}\gamma^{*}+\left(|f_{2}|^{2}g_{2}^{*}g_{3}+f_{2}^{*}f_{1}g_{2}^{*}g_{4}\right)\beta^{*2}\gamma\right.\\
 & + & \left.\left(|g_{1}|^{2}-|g_{2}|^{2}\right)\left\{ \left(f_{4}^{*}f_{2}-f_{3}^{*}f_{1}\right)\alpha\beta\gamma^{*}-\left(g_{2}^{*}g_{4}-g_{1}^{*}g_{3}\right)\alpha^{*}\beta^{*}\gamma\right\} \right].\end{array}\label{eq:hz2-ab1}\end{equation}
It is easy to observe that Eqns. (\ref{eq:HZ-1-ab1}) and (\ref{eq:hz2-ab1})
provide us the following simple relation that is valid for the present
case: $E_{ab_{1}}^{1,1}=-E_{ab_{1}}^{\prime1,1}$, which implies that
for any particular choice of rescaled interaction length $\Gamma L$
either HZ-I criterion or HZ-II criterion would show the existence
of entanglement in contradirectional asymmetric nonlinear optical
coupler as both of them cannot be simultaneously positive. Thus the
compound mode $ab_{1}$ is always entangled. The same is explicitly
illustrated through Fig. \ref{fig:ent-hz1-II}. Similar investigations
using HZ-I and HZ-II criteria in the other two compound modes (i.e.,
$ab_{2}$ and $b_{1}b_{2})$ failed to obtain any signature of entanglement
in these cases. Further, signature of intermodal entanglement was
not witnessed using Duan et al. criterion as using the present solution
and (\ref{duan-1}) we obtain \begin{equation}
d_{ab_{1}}=d_{ab_{2}}=d_{b_{1}b_{2}}=0.\label{eq:variance-duan}\end{equation}
 However, it does not ensure separability of these modes as HZ-I,
HZ-II and Duan et al. inseparability criteria are only sufficient
and not necessary. 

\begin{figure}
\begin{centering}
\includegraphics[scale=0.5]{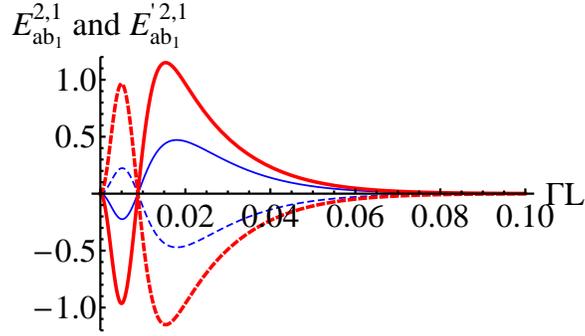}
\par\end{centering}

\caption{{\large \label{fig:hoe}}(color online) Higher order entanglement
is observed using Hillery-Zubairy criteria. Solid lines show spatial
variation of $E_{ab_{1}}^{2,1}$ and dashed lines show spatial variation
of $E_{ab_{1}}^{\prime2,1}$ with $k=0.1,\,\Gamma=0.001,\,\Delta k=10^{-4},\,\beta=2,\gamma=1$
for $\alpha=3$ (thin blue lines) and $\alpha=5$ (thick red lines).
It is observed that depth of nonclassicality increases with increase
in $\alpha$.}
\end{figure}

We may now study the possibilities of existence of higher order entanglement
using Eqns. (\ref{hoe-criteria-1})-(\ref{eq:fully ent2}). To begin
with, we use (\ref{2})-(\ref{eq:soln}) and (\ref{hoe-criteria-1})
to yield \begin{equation}
\begin{array}{lcl}
E_{ab_{1}}^{m,n} & = & \langle a^{\dagger m}a^{m}b_{1}^{\dagger n}b_{1}^{n}\rangle-|\langle a^{m}b_{1}^{\dagger n}\rangle|^{2}\\
 & = & mn\left|\left(f_{1}\alpha+f_{2}\beta\right)\right|^{2m-2}\left|\left(g_{1}\alpha+g_{2}\beta\right)\right|^{2n-2}E_{ab_{1}}^{1,1}.\end{array}\label{eq:emnab1}\end{equation}
Similarly, using (\ref{2})-(\ref{eq:soln}) and (\ref{hoe-criteria-2})
we can produce an analytic expression for $E_{ab_{1}}^{\prime m,n}$
and observe that \begin{equation}
E_{ab_{1}}^{\prime m,n}=-E_{ab_{1}}^{m,n}.\label{eq:emnprimeab1}\end{equation}
From relation (\ref{eq:emnprimeab1}), it is clear that the higher
order entanglement between $a$ mode and $b_{1}$ mode would always
exist for any choice of $\Gamma L,$ $m$ and $n$. This is so because
$E_{ab_{1}}^{m,n}$ and $E_{ab_{1}}^{\prime m,n}$ cannot be simultaneously
positive. Using (\ref{eq:emnab1}) and (\ref{eq:emnprimeab1}), it
is a straight forward exercise to obtain analytic expressions of $E_{ab_{1}}^{m,n}$
and $E_{ab_{1}}^{\prime m,n}$ for specific values of $m$ and $n.$
Such analytic expressions are not reported here as the existence of
higher order entanglement is clearly observed through (\ref{eq:emnprimeab1}).
However, in Fig. \ref{fig:hoe} we have illustrated the variation
of $E_{ab_{1}}^{2,1}$ and $E_{ab_{1}}^{\prime2,1}$ with the rescaled
interaction length, $\Gamma L$. Negative parts of the plots shown
in Fig. \ref{fig:hoe} illustrate the existence of higher order intermodal
entanglement in compound mode $ab_{1}.$ As expected from (\ref{eq:emnprimeab1}),
we observe that for any value of $\Gamma L$ compound mode $ab_{1}$
is higher order entangled. Further, it is observed that Hillery-Zubairy's
higher order entanglement criteria (\ref{hoe-criteria-1})-(\ref{hoe-criteria-2})
fail to detect any signature of higher order entanglement in compound
modes $ab_{2}$ and $b_{1}b_{2}.$

One can also investigate higher order entanglement using criterion
of multi-partite (multi-mode) entanglement as all the multi-mode entangled
states are essentially higher order entangled. Here we have only three
modes in the coupler and thus we can study higher order entanglement
by investigating the existence of three-mode entanglement. A three-mode
pure state that violates (\ref{eq:fully ent2}) (i.e., satisfies $\langle N_{a}\rangle\langle N_{b_{1}}\rangle\langle N_{b_{2}}\rangle-|\langle ab_{1}b_{2}\rangle|^{2}<0)$
and simultaneously satisfies either (\ref{eq:fully enta 0}) or (\ref{eq:fully enta 1})
is a fully entangled state. Using (\ref{2})-(\ref{eq:soln}) and
(\ref{eq:tripartite ent1})-(\ref{eq:fully ent2}) we obtain following
set of interesting relations for $m=n=l=1$: 

\begin{equation}
E_{a|b_{1}b_{2}}^{1,1,1}=-E_{a|b_{1}b_{2}}^{\prime1,1,1}=E_{ab_{2}|b_{1}}^{1,1,1}=-E_{ab_{2}|b_{1}}^{\prime1,1,1}=|\gamma|^{2}E_{ab_{1}}^{1,1},\label{eq:relation1}\end{equation}
 \begin{equation}
E_{ab_{1}|b_{2}}^{1,1,1}=E_{ab_{1}|b_{2}}^{\prime1,1,1}=0,\label{eq:relation2}\end{equation}
and \begin{equation}
\langle N_{a}\rangle\langle N_{b_{1}}\rangle\langle N_{b_{2}}\rangle-|\langle ab_{1}b_{2}\rangle|^{2}=-|\gamma|^{2}E_{ab_{1}}^{1,1}.\label{eq:relation3}\end{equation}
From (\ref{eq:relation1}), it is easy to observe that three modes
of the coupler are not bi-separable in the form $a|b_{1}b_{2}$ and
$ab_{2}|b_{1}$ for any choice of rescaled interaction length $\Gamma L>0.$
Further, Eqn. (\ref{eq:relation3}) shows that the three modes of
the coupler are not fully separable for $E_{ab_{1}}^{1,1}>0$ (c.f.
positive regions of plot of $E_{ab_{1}}^{1,1}$ shown in Fig. \ref{fig:ent-hz1-II}).
However, (\ref{eq:relation2}) illustrate that the present solution
does not show entanglement between coupled mode $ab_{1}$ and single-mode
$b_{2}.$ Thus the three modes present here are not found to be fully
entangled. Specifically, three-mode (higher order) entanglement is
observed here, but signature of fully entangled three-mode state is
not observed. Further, we have observed that in all the figures depth
of nonclassicality increases with $\alpha.$

In the present paper, various lower order and higher order nonclassical
phenomena have been observed in contradirectional asymmetric nonlinear
optical coupler. However, so far we have discussed only the stimulated
cases as no nonclassical phenomenon is expected to be observed in
spontaneous case. This is so because all the useful non-vanishing
expressions for witnessing nonclassicality (i.e., Eqs. (\ref{eq:asq-1})-(\ref{eq:relation3}))
are proportional to $|\gamma|.$ Thus all these expressions would
vanish for $\gamma=0$. 

\begin{table}
\begin{centering}
\begin{tabular}{|c|>{\centering}p{2in}|c|c|c|}
\hline 
S.No. & Nonclassical phenomenon & Modes & Short-length approximation \cite{contra-1,co and contra with phase mismatch} & Present work\tabularnewline
\hline 
1. & Squeezing  & $b_{2}$ & Not investigated & Not observed\tabularnewline
\hline 
2. & Intermodal squeezing  & $b_{1}b_{2},\, ab_{2}$ & Not observed & Observed\tabularnewline
\hline 
3. & Amplitude squared squeezing  & $a,\, b_{1}$ & Not investigated & Observed\tabularnewline
\hline 
4. & Amplitude squared squeezing  & $b_{2}$ & Not investigated & Not observed\tabularnewline
\hline 
5. & Lower order and higher order intermodal entanglement & $ab_{1}$ & Not investigated & Observed\tabularnewline
\hline 
6. & Lower order and higher order intermodal entanglement  & $b_{1}b_{2},\, ab_{2}$ & Not investigated & Not observed\tabularnewline
\hline 
7. & Three-mode (higher order) bi-separable entanglement & $a|b_{1}b_{2}$, $ab_{2}|b_{1}$ & Not investigated & Observed\tabularnewline
\hline 
8. & Three-mode (higher order) bi-separable entanglement & $a|b_{1}b_{2}$ & Not investigated & Not observed\tabularnewline
\hline
\end{tabular}
\par\end{centering}

\caption{\label{tab:Nonclassicalities-observed-in}Nonclassicalities observed
in a contradirectional asymmetric nonlinear optical coupler that were
not observed in earlier studies \cite{contra-1,co and contra with phase mismatch}.}
\end{table}

\section{{\normalsize Conclusions\label{sec:Conclusions}}}

In the present study we report lower order and higher order nonclassicalities
in a contradirectional asymmetric nonlinear optical coupler using
a set of criteria of entanglement, single-mode squeezing, intermodal
squeezing, antibunching, intermodal antibunching etc. Variation of
nonclassicality with various parameters, such as number of input photon
in the linear mode, linear coupling constant, nonlinear coupling constant
and phase mismatch is also studied and it is observed that amount
of nonclassicality can be controlled by controlling these parameters.
The contradirectional asymmetric nonlinear optical coupler studied
in the present work was studied earlier using a short-length solution
and considering $b_{2}$ mode as classical \cite{contra-1,co and contra with phase mismatch}.
In contrast, a completely quantum mechanical solution of the equations
of motion is obtained here using Sen-Mandal approach which is not
restricted by length. The use of better solution and completely quantum
mechanical treatment led to the identification of several nonclassical
characters of a contradirectional asymmetric nonlinear optical coupler
that were not reported in earlier studies. All such nonclassical phenomena
that are observed here and were not observed in earlier studies are
listed in Table \ref{tab:Nonclassicalities-observed-in}. Further,
there exist a large number of nonclassicality criteria that are not
studied here and are based on expectation values of moments of annihilation
and creation operators (c.f. Table I and II of Ref. \cite{Adam-criterion}).
As we already have compact expressions for the field operators it
is a straight forward exercise to extend the present work to investigate
other signatures of nonclassicality, such as, photon hyperbunching
\cite{hyperbunching}, sum and difference squeezing of An-Tinh \cite{sum&difference-sq-An-Tinh}
and Hillery \cite{Sum&diff-sq-Hillery}, inseparability criterion
of Manicini et al. \cite{Mancini-ent}, Simon \cite{Simon} and Miranowicz
et al. \cite{Adam-ent} etc. Further, the present work can be extended
to investigate the nonclassical phenomena in other types of contradirectional
optical couplers (e.g., contradirectional parametric coupler, contradirectional
Raman coupler, etc.) that are either not studied till date or studied
using short-length solution. Recently, Allevi et al. \cite{Maria-PRA-1,Maria-2}
and Avenhaus et al. \cite{higher-order-PRL} have independently reported
that they have experimentally measured $\langle a_{1}^{\dagger j}a_{1}^{j}a_{2}^{\dagger k}a_{2}^{k}\rangle$
which is sufficient to completely characterize bipartite multi-mode
states. It is easy to observe that ability to experimentally measure
$\langle a_{1}^{\dagger j}a_{1}^{j}a_{2}^{\dagger k}a_{2}^{k}\rangle$
ensures that we can experimentally detect signatures of nonclassicalities
reported here. It is true that most of the nonclassicalities reported
here can be observed in some other bosonic systems, too. However,
the present system has some intrinsic advantages over most of the
other systems as it can be used as a component in the integrated waveguide
based structures in general and photonic circuits in particular {[}
\cite{Jasleen-thesis,Integrated Quant. Opti} and references therein{]}.
Thus the nonclassicalities reported in this easily implementable waveguide
based system is expected to be observed experimentally. Further, the
system studied here is expected to play important role as a source
of nonclassical fields in the integrated waveguide based structures.

\textbf{Acknowledgment:} K. T. and A. P. thank the Department of Science
and Technology (DST), India, for support provided through DST project
No. SR/S2/LOP-0012/2010 and A. P. also thanks the Operational Program
Education for Competitiveness-European Social Fund project CZ.1.07/2.3.00/20.0017
of the Ministry of Education, Youth and Sports of the Czech Republic.
J. P. thanks the Operational Program Research and Development for
Innovations - European Regional Development Fund project CZ.1.05/2.1.00/03.0058
of the Ministry of Education, Youth and Sports of the Czech Republic.


\begin{thebibliography}{71}
\bibitem{CV-qkd-hillery}M. Hillery, Phys. Rev. A \textbf{61} (2000)
022309.

\bibitem{teleportation of coherent state}A. Furusawa, J. L. Sorensen,
S. L. Braunstein, C. A. Fuchs, H. J. Kimble and E. S. Polzik, Science
\textbf{282} (1998) 706.

\bibitem{antibunching-sps}Z. Yuan, B. E. Kardynal, R. M. Stevenson,
A. J. Shields, C. J. Lobo, K. Cooper, N. S. Beattie, D. A. Ritchie
and M. Pepper, Science \textbf{295} (2002) 102.

\bibitem{Ekert protocol}A. Ekert, Phys. Rev. Lett. \textbf{67} (1991)
661.

\bibitem{Bennet1993}C. H. Bennett, G. Brassard, C. Crepeau, R. Jozsa,
A. Peres and W. K. Wootters , Phys. Rev. Lett. \textbf{70 } (1993)
1895.

\bibitem{densecoding}C. H. Bennett and S. J. Wiesner, Phys. Rev.
Lett. \textbf{69} (1992) 2881.

\bibitem{device ind}A. Acin, N. Gisin and L. Masanes, Phys. Rev.
Lett. \textbf{97} (2006) 120405.

\bibitem{pathak-perina}A. Pathak, J. K\v{r}epelka and Jan ${\rm Pe\check{r}ina}$,
Phys. Lett. A \textbf{377} (2013) 2692.

\bibitem{pathak-PRA}B. Sen, S. K. Giri, S. Mandal, C. H. R. Ooi and
A. Pathak, Phys. Rev. A \textbf{87} (2013) 022325.

\bibitem{pathak-pra2}S. K. Giri, B. Sen, C. H. R. Ooi and A. Pathak,
Phys. Rev. A\textbf{ 89} (2014) 033628.

\bibitem{Nature09} J. C. F. Matthews, A. Politi, A. Stefanov and
J. L. O\textquoteright{}Brien, Nature Photonics, \textbf{3} (2009)
346.

\bibitem{Mandal-Midda} P. Mandal and S. Midda, Optik \textbf{122}
(2011) 1795.

\bibitem{kishore-1}K. Thapliyal, A. Pathak, B. Sen and J. J. ${\rm Pe\breve{r}ina}$,
arXiv:1403.6647 (2014).

\bibitem{perina-review} J. ${\rm Pe\breve{r}ina}$ Jr. and J. ${\rm Pe\breve{r}ina}$,
in: E. Wolf (Ed.), Progress in Optics, vol. 41, Elsevier, Amsterdam,
(2000), 361. 

\bibitem{contra-1} J. ${\rm Pe\breve{r}ina}$ and J. ${\rm Pe\breve{r}ina}$
Jr., Quantum Semiclass. Opt. \textbf{7} (1995) 849.

\bibitem{bsen1} B Sen and S Mandal, J. Mod. Opt. \textbf{52} (2005)
1789.

\bibitem{Bsen-2}B. Sen, S. Mandal and J. ${\rm Pe\breve{r}ina}$
, J. Phys. B: At. Mol. Opt. Phys. \textbf{40} (2007) 1417.

\bibitem{bsen3} B Sen and S Mandal, J. Mod. Opt. \textbf{55} (2008)
1697.

\bibitem{Mandal-Perina}S. Mandal and J. ${\rm Pe\breve{r}ina}$,
Phys. Lett. A \textbf{328} (2004) 144. 

\bibitem{Kerr-1}F. A. A. El-Orany, M. Sebawe Abdalla and J. ${\rm Pe\breve{r}ina}$,
Eur. Phys. J. D, \textbf{33} (2005) 453.

\bibitem{kerr-2}N. Korolkova and J. ${\rm Pe\breve{r}ina}$, Opt.
Commun. \textbf{136} (1997) 135. 

\bibitem{Kerr-3}J. Fiurasek, J. K\v{r}epelka and J. ${\rm Pe\breve{r}ina}$,
Opt. Commun. \textbf{167} (1999) 115. 

\bibitem{kerr-4}G. Ariunbold and J. ${\rm Pe\breve{r}ina}$, Opt.
Commun. \textbf{176} (2000) 149.

\bibitem{kerr-5}N. Korolkova and J. ${\rm Pe\breve{r}ina}$, J. Mod.
Opt.\textbf{ 44} (1997) 1525. 

\bibitem{Raman and Brilloin} J. ${\rm Pe\breve{r}ina}$ Jr. and J.
${\rm Pe\breve{r}ina}$, Quantum Semiclass. Opt. \textbf{9} (1997)
443.

\bibitem{parametric}N. Korolkova and J. ${\rm Pe\breve{r}ina}$,
Opt. Commun. \textbf{137} (1997) 263. 

\bibitem{parametric2}J. ${\rm Pe\breve{r}ina}$ and J. ${\rm K\check{r}epelka}$,
Opt. Commun. \textbf{326} (2014) 10.

\bibitem{assymetric-strong-pump}J. Perina and J. ${\rm Pe\breve{r}ina}$
Jr., Quantum Semiclass. Opt. \textbf{7} (1995) 541.

\bibitem{assymetric-2}J. ${\rm Pe\breve{r}ina}$, J. Mod. Opt. \textbf{42}
(1995) 1517. 

\bibitem{nonclassical-input1}J. ${\rm Pe\breve{r}ina}$ and J. Bajer,
J. Mod. Opt. \textbf{42} (1995) 2337. 

\bibitem{co-and-contra}J. ${\rm Pe\breve{r}ina}$ and J. ${\rm Pe\breve{r}ina}$
Jr., J. Mod. Opt. \textbf{43} (1996) 1951.

\bibitem{nonclasssical input2}L. Mi\v{s}ta Jr. and J. ${\rm Pe\breve{r}ina}$,
Czechoslovak Journal of Physics, \textbf{47} (1997) 629.

\bibitem{co and contra with phase mismatch} J. ${\rm Pe\breve{r}ina}$
and J. ${\rm Pe\breve{r}ina}$ Jr., Quantum Semiclass. Opt. \textbf{7}
(1995) 863.

\bibitem{leonoski1} A. Kowalewska-Kud\l{}aszyk and W. Leonski, J.
Opt. Soc. B \textbf{26} (2009) 1289.

\bibitem{thermal ent-kerr} M.R. Abbasi and M.M. Golshan, Physica
A \textbf{392} (2013) 6161.

\bibitem{kerr-lionoski-miranowicz}W. Leonski and A. Miranowicz, J.
Opt. B: Quantum Semiclass. Opt. \textbf{6} (2004) S37.

\bibitem{GBS} A. Kowalewska-Kud\l{}aszyk, W. Leonski and J. ${\rm Pe\breve{r}ina}$
Jr., Phys. Scr. \textbf{T147} (2012) 014016.

\bibitem{higher-order} F. A. A. El-Orany and J. ${\rm Pe\breve{r}ina}$,
Phys. Lett. A \textbf{333} (2004) 204.

\bibitem{adam}A. Miranowicz and W. Leonski, J. Phys. B \textbf{39}
(2006) 1683.

\bibitem{Loudon} R. Loudon and P. L. Night, J. Mod. Opt. \textbf{34}
(1987) 709.

\bibitem{NJP-GS-Ashoka} G. S. Agarwal and A. Biswas, New J. Phys.
\textbf{7} (2005) 211.

\bibitem{Adam-criterion}A. Miranowicz, M. Bartkowiak, X. Wang, Y.-x.
Liu and F. Nori, Phys. Rev. A \textbf{82} (2010) 013824.

\bibitem{HZ-PRL} M. Hillery and M. S. Zubairy, Phys. Rev. Lett. \textbf{96}
(2006) 050503.

\bibitem{HZ2007} M. Hillery and M. S. Zubairy, Phys. Rev. A \textbf{74}
(2006) 032333.

\bibitem{HZ2010} M. Hillery, H. T. Dung and H. Zheng, Phys. Rev.
A \textbf{81} (2010) 062322.

\bibitem{duan} L. M. Duan, G. Giedke, J. I. Cirac and P. Zoller,
Phys. Rev. Lett. \textbf{84} (2000) 2722.

\bibitem{vogel's ent criterion}E. Shchukin and W. Vogel, Phys. Rev.
Lett. \textbf{95} (2005) 230502. 

\bibitem{Adam-generalizes voge;}A. Miranowicz, M. Piani, P. Horodecki
and R. Horodecki, Phys. Rev. A \textbf{80} (2009) 052303.

\bibitem{Perina-Book}J. ${\rm Pe\breve{r}ina}$, Quantum Statistics
of Linear and Nonlinear Optical Phenomena, Kluwer, Dordrecht (1991).

\bibitem{C T Lee}C. T. Lee, Phys. Rev. A \textbf{41} (1990) 1721.

\bibitem{HOAis not rare}P. Gupta, P. Pandey and A. Pathak, J. Phys.
B \textbf{39 }(2006) 1137 .

\bibitem{generalized-higher order}A. Verma and A. Pathak, Phys. Lett.
A \textbf{374} (2010) 1009.

\bibitem{HOAwithMartin}A. Pathak and M. Garcia, Applied Physics B
\textbf{84} (2006) 484.

\bibitem{HIlery-amp-sq}M. Hillery, Phys. Rev. A \textbf{36} (1987)
3796.

\bibitem{Mancini-ent}S. Mancini, V. Giovannetti, D. Vitali, and P.
Tombesi, Phys. Rev. Lett. \textbf{88} (2002) 120401.

\bibitem{Simon}R. Simon, Phys. Rev. Lett. \textbf{84} (2000) 2726. 

\bibitem{Adam-ent}A. Miranowicz, M. Piani, P. Horodecki, and R. Horodecki,
Phys. Rev. A \textbf{80} (2009) 052303 .

\bibitem{Hong-Mandel1}C. K. Hong and L. Mandel, Phys. Rev. Lett.
\textbf{54} (1985) 323. 

\bibitem{HOng-mandel2}C. K. Hong and L. Mandel, Phys. Rev. A \textbf{32}
(1985) 974. 

\bibitem{higher order multiparty1} A. Zeilinger, M. A. Horne and
D. M. Greenberger, NASA Conf. Publ. \textbf{3135} (1992) 51.

\bibitem{higher order multyparty2}J.-W. Pan, M. Daniell, S. Gasparoni,
G. Weihs and A. Zeilinger, Phys. Rev. Lett. \textbf{86} (2001) 4435.

\bibitem{higher order multyparty3}A. Mair, A. Vaziri, G. Weihs and
A. Zeilinger, Nature \textbf{412} (2001) 313.

\bibitem{Ent condition-multimode}Z.-G. Li, S.-M. Fei, Z.-X. Wang
and K. Wu, Phys. Rev. A \textbf{75} (2007) 012311.

\bibitem{hyperbunching} M. Jakob, Y. Abranyos, and J. A. Bergou,
J. Opt. B: Quantum Semiclass. Opt. \textbf{3} (2001) 130.

\bibitem{sum&difference-sq-An-Tinh}N. B. An and V. Tinh, Phys. Lett.
A \textbf{261} (1999) 34; N. B. An and V. Tinh, Phys. Lett. A \textbf{270}
(2000) 27.

\bibitem{Sum&diff-sq-Hillery}M. Hillery, Phys. Rev. A \textbf{40},
3147 (1989).

\bibitem{Maria-PRA-1}A. Allevi, S. Olivares and M. Bondani, Phys.
Rev. A \textbf{85} (2012) 063835.

\bibitem{Maria-2}A. Allevi, S. Olivares and M. Bondani, Int. J. Quant.
Info. \textbf{8} (2012) 1241003.

\bibitem{higher-order-PRL}M. Avenhaus, K. Laiho, M. V. Chekhova and
C. Silberhorn, Phys. Rev. Lett \textbf{104} (2010) 063602.

\bibitem{Jasleen-thesis}J. Lugani, \textquotedbl{}Studies on guided
wave devices and ultracold atoms in optical cavity for applications
in quantum optics and quantum information'', PhD thesis, IIT Delhi
(2013), http://web.iitd.ac.in/\textasciitilde{}sankalpa/Jasleen\_corrected\_thesis.pdf.

\bibitem{Integrated Quant. Opti}S. Tanzilli, A. Martin, F. Kaiser,
M. P. De Micheli, O. Alibart and D. B. Ostrowsky, Laser \& Photonics
Rev. \textbf{6} (2012) 115.
\end{thebibliography}
\end{document}